%% file: 3emom.tex
\documentclass[aps,prl,reprint,twocolumn,showpacs,floatfix,superscriptaddress]{revtex4-1}

\usepackage{amssymb,amsmath,amstext}                

\usepackage{graphicx}
\usepackage[usenames, dvipsnames]{color}
\usepackage{dcolumn}
\usepackage{bm}
\usepackage{color}
\usepackage[colorlinks,bookmarks=false,citecolor=blue,linkcolor=blue,urlcolor=black]{hyperref}
\usepackage{ulem}

\usepackage{comment}

\begin{document}

\preprint{APS/123-QED}


\title{Three-electron correlations in strong laser field ionization: Spin induced effects} 
\author{Dmitry K. Efimov}
\affiliation{Institute of Theoretical Physics, Jagiellonian University in Krakow, Łojasiewicza 11, 30-348 Kraków, Poland}
\affiliation{Department of Theoretical Physics, Faculty of Fundamental Problems of Technology,
Wrocław University of Science and Technology, 50-370 Wrocław, Poland}

\author{Artur Maksymov}
\affiliation{Institute of Theoretical Physics, Jagiellonian University in Krakow, Łojasiewicza 11, 30-348 Kraków, Poland}

\author{Marcelo Ciappina}
\affiliation{ICFO - Institut de Ciencies Fotoniques, The Barcelona Institute of Science and Technology, \\ Av. Carl Friedrich Gauss 3,
	08860 Castelldefels (Barcelona), Spain}
\affiliation{Physics Program, Guangdong Technion–Israel Institute of Technology, Shantou 515063, China}
\affiliation{Technion–Israel Institute of Technology, Haifa 32000, Israel}

\author{ Jakub S. Prauzner-Bechcicki}
\affiliation{Instytut Fizyki imienia Mariana Smoluchowskiego, Uniwersytet Jagiello\'nski, \L{}ojasiewicza 11, 30-348 Krak\'ow, Poland}

\author{Maciej Lewenstein}
\affiliation{ICFO - Institut de Ciencies Fotoniques, The Barcelona Institute of Science and Technology, \\ Av. Carl Friedrich Gauss 3,
	08860 Castelldefels (Barcelona), Spain}
\affiliation{ICREA, Pg. Lluís Companys 23, 08010 Barcelona, Spain}

\author{Jakub Zakrzewski}
\affiliation{Institute of Theoretical Physics, Jagiellonian University in Krakow, Łojasiewicza 11, 30-348 Kraków, Poland}
\affiliation{ Mark Kac Complex Systems Research Center, Jagiellonian University, \L{}ojasiewicza 11, 30-348 Krak\'ow, Poland}
 \email{jakub.zakrzewski@uj.edu.pl}

\date{\today}

\begin{abstract}
Strong field processes in the non-relativistic regime are insensitive to the electron spin, i.e. the observables appear to be independent of this electron property. This does not have to be the case for several active electrons where Pauli principle may affect the their dynamics.
We exemplify this statement studying model atoms with three active electrons interacting with strong pulsed radiation, using an ab-initio time-dependent Schr\"odinger 
equation on a grid. In our restricted dimensionality model we are able, for the first time, to analyse momenta correlations of the three outgoing 
electrons using Dalitz plots. We show that significant differences are obtained between model Neon and Nitrogen atoms. 
These differences are traced back to the different symmetries of the electronic wavefunctions, and directly related to the different initial state spin components.
\end{abstract}

\maketitle


	\noindent{\it Introduction:} Interaction of strong laser light with matter still brings new surprises for more than half century as reviewed by  \cite{Krausz09}. It was realized quite early that the impact of short but strong laser radiation may lead to multiple ionization of atoms 
 \cite{Lhuillier82,Lhuillier83,Luk83,Boyer84,Luk85,Chin85} and subsequently discussion on collective, non-sequential or sequential character of electrons emission has begun \cite{Lhuillier83,Lambropoulos85,Yergeau87,Crance87,Mu86,Aberg84,Geltman85,Zakrzewski86,Lewenstein86,Geltman88}.
It soon became apparent that the physics involved seems to be more interesting at intermediate intensities when the existence of the characteristic knee in the ionization yield triggered a lively discussion about the importance of the so-called re-scattering mechanism on the non-sequential double ionization~\cite{Fittinghoff92, Kondo93, walker1994precision}. The theoretical picture came with the simple-man three-step model
\cite{Corkum93}, that provided a simple and intuitively explanation of above-threshold ionization (with electrons leaving the atom with energy far exceeding the threshold \cite{Eberly91}) as well as high harmonic generation (HHG) \cite{Lewenstein94} and re-scattering mechanism leading to enhanced double ionization 
\cite{Fittinghoff92, Kondo93, walker1994precision}.  
The cold-target recoil-ion-momentum spectroscopy (COLTRIMS)  \cite{dorner00,weber00,Moshammer2000-vn,Liu2008-yt,Kubel19} allowed one to detect the momenta of an ion and ejected electrons during the ionization event opening the path for the detailed study of  both the electron and ion laser-induced  dynamics. Further in-depth studies of electron-electron correlation during strong-field evolution led to deeper understanding of ongoing processes and the development of more sophisticated models \cite{Chen10,Chen19,Maxwell16}. For an additional resolution of ionization delays, the coincidence scheme was assisted by a streaking technique \cite{Winney17,Winney18,Kubel19}. An alternative path for accessing real-time dynamics implies a combination of coincidence spectroscopy with an attosecond interferometry \cite{Zhong20,Mikaelsson20}. 
	
	Importantly, recent coincidence experimental schemes \cite{Bergues12,Henrichs18,Larimian20,Grundmann20,Zhong20,Mikaelsson20} are kinematically complete, that is, momenta, position and detection time are measured for each charged particle explicitly. With increased  energy resolution \cite{Zhong20,Mikaelsson20} one is able to catch fine interference structures that form
the quantum fingerprint of the laser-driven electron dynamics. Those recent advances removed the limitations on the number of particles detected: there is no need need to use ionic momenta for recalculating electronic ones.
		
	{Although challenging,} the experimental arena is thus ready for three-electron coincidence experiments. Specially considering that, in ion-impact induced ionization, 
	the pioneering experiments were already done long time ago \cite{Schulz00}. Such an experiment for strong field ionization would definitely open a new page in the understanding of electronic correlations. Its power is of particular importance for studies of recollisional dynamics induced by IR femtosecond laser pulses. In particular, different double ionization channels of multi-electron atoms (affecting different electrons) cannot be resolved in a two-electron coincidence experiment\cite{Efimov19}; the three-electron coincidence scheme, on the other hand, potentially could allow one to track the relative contributions of different channels \cite{Prauzner-Bechcicki21}.

{The theoretical support needed for these prospective experiments are, however, 
not still available partly because numerical simulations for many electron ionization 
are extremely challenging. A full quantum-mechanical simulations are, at best, 
possible for two electrons thanks to the efforts of Ken Taylor's group in Belfast~\cite{Parker98,Parker00}. 
Furthermore, models for three electrons are at their infancy 
being limited to either purely classical Monte Carlo methods \cite{Emmanouilidou08b,Peters20,ho2006plane,ho2007argon,Guo08,Yuan19}
 or 
to time-dependent orbital approaches which, 
while successful for description of HHG\cite{Sato16},
can hardly provide detailed information on the ionization 
dynamics and momenta distributions.}

	Recently, we presented introductory studies of strong-field triple ionization, however, due to the intrinsic complexity of the problem, inevitably we restricted ourselves to only investigate the dependence of ionization yields on the laser pulse amplitude~\cite{Thiede18,Efimov19,Efimov20,Prauzner-Bechcicki21}. In the present work, we raise the analysis to a significantly higher level, addressing three electrons momenta distributions. Our method is based on the Eckhardt-Sacha model of reduced dimensionality, derived from the analysis of saddles in the effective adiabatic potential for electrons in an instantaneous electric field~\cite{Sacha01}. Let us emphasize, that this is a distinctly different approach from the standard Rochester reduction, which restrain the motion of electrons to the laser polarization axis~\cite{Grobe92,Liu99}.

\noindent{\it The model:} The Hamiltonian of the model studied reads \cite{Thiede18}: 
\begin{equation}
H=\sum_{i=1}^3\frac{p_i^2}{2}+V(r_1,r_2,r_3)
\label{ham3e}
\end{equation}
with 
\begin{eqnarray}
V(r_1,r_2,r_3)&=&-\sum_{i=1}^3\left(\frac{3}{\sqrt{r_i^2+\epsilon^2}} +\sqrt{\frac{2}{3}}F(t)r_i \right) \nonumber \\
&+&\sum_{i,j=1 i<j}^3\frac{q_{ee}^2}{\sqrt{(r_i-r_j)^2+r_ir_j+\epsilon^2}},
\label{pot3e_std}
\end{eqnarray}
%
where $r_i$ and $p_i$ correspond to the $i$-th electron's position and momentum, respectively and $\epsilon$ is a parameter softening the Coulomb singularity. The driving time dependent field
$F(t) = -\partial A/\partial t$ is given via the vector potential
\begin{equation}
A(t) = \frac{F_0}{\omega_0} \sin^2 \left( \frac{\pi t}{T_p} \right) \sin(\omega_0 t + \phi), \quad 0<t<T_p,
\label{pulse}
\end{equation}
with the pulse length $T_p = 2\pi n_c /\omega_0$ {and carrier-envelope phase (CEP) $\phi$ which we put to zero}. We limit here to $n_c=3$ optical cycles case and assume $\omega_0=0.06$ a.u., that corresponds to a laser wavelength of about 760 nm. The smoothing parameter $\epsilon=\sqrt{0.83}$ a.u. {and effective electric charge $q_{ee}=1$} assures that
 the  ground state energy is equal to the triple ionization potential of Neon, 
$I_p=4.63$ a.u., i.e. 126 eV. (for appropriate symmetry of the wavefunction, see below). 
Starting from the ground state, found with imaginary time propagation method,
we directly solve the time-dependent Schr\"odinger equation (TDSE) on a large 3-dimensional grid using a standard propagation scheme \cite{Thiede18}. The algorithm used for simulating momentum distributions is a direct extension of the approach introduced for the two-electron case \cite{Lein00,Prauzner08}. For the sake of brevity, will describe it very shortly:  for each electronic coordinate we distinguish the ``bounded motion'' region where the numerical solution of the TDSE is used and the ``outer" region where each electron is assumed to move freely and the interactions with other electrons and the nucleus are neglected. A distinction between "bounded motion" and "outer" regions is done by setting the threshold distance, {defined by a classical distance of no return, $r_t=F_0/\omega_0^2$} from the center of the coordinate system -- when a given electron coordinate value exceeds $r_t$ it is evolved in the "outer" region; the electron never comes back to the "bounded motion" region. The solutions from the different regions are added coherently (including the possible probability fluxes between the regions). The conceptual idea  was created in \cite{Lein00} for two-electron problems (see also \cite{Prauzner08} for a detailed description) and is generalized here to the three-electron case.

\begin{figure}[t]
	\includegraphics[width=0.5\linewidth]{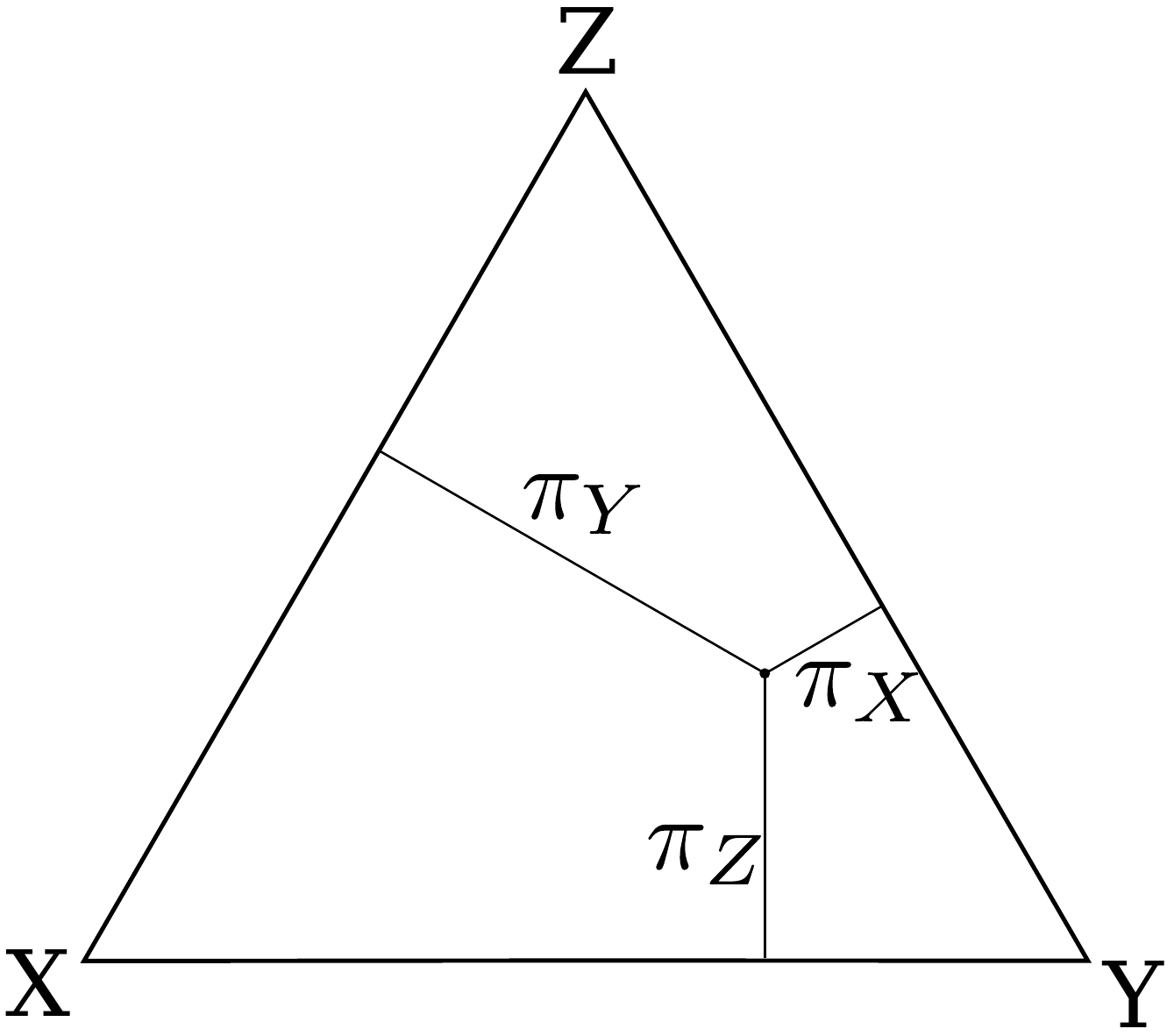}
	\caption{ Structure of a Dalitz plot. The three-electron ($X,Y,Z$) momentum distribution is mapped into a triangle. Each point inside the triangle is defined by coordinates $\{ \pi_X,\pi_Y,\pi_Z \}$. For each electron $i$, the small $\pi _i$ the small is the ratio of the momentum $p_i$ to the triple-ion momentum value.
 See the text for more details. }
	\label{fig1}
\end{figure}
Additionally, and novel for three electrons, a twist comes from the fact that for three electrons one has to be careful about the spatial symmetries of the wavefunction. As is known they are directly related to electronic spin and Pauli exclusion principle. 
This has been already noticed in the pioneering studies of Li atom ionization \cite{Ruiz05,Ruiz06}. In its $1s^22s^1$ initial state the corresponding Slater wavefunction is a sum of 3 terms (each corresponding to electron permutations) with two electrons with spin up (U) and one down (D).
Since the Hamiltonian is spin-independent (neglecting spin-orbit coupling) for non-relativistic strong laser-atom interactions all three components evolve identically and a single combination may be time evolved only. The same would occur for species with electronic $ns^2np^1$ configuration, e.g. for Boron or Aluminum, however, in these cases a real multi-photon regime falls within very low frequencies. As a  model system with electronic $ns^2np^1$ configuration we consider an artificial three-electron atom with the first three ionization thresholds corresponding to Neon \cite{Efimov19}. Thus, we will refer to that model as Neon. The ground state wavefunction in the model atom is spatially partially antisymmetric  (assuming  the spin configuration is (UUD) meaning Up-Up-Down for X, Y and Z electrons): $\Psi(X,Y,Z)=\Psi(X,Z,Y)=-\Psi(Y,X,Z)=-\Psi(Z,Y,X)$ \cite{Thiede18}.  For comparison, we will consider a model atom that has $p^3$ configuration with all spins oriented in the same direction (say UUU) and, thus, a totally antisymmetric spatial wavefunction : $\Psi(X,Y,Z)=-\Psi(X,Z,Y)=-\Psi(Y,X,Z)=-\Psi(Z,Y,X)$. The corresponding ionization potentials are set to match the first three ionization potentials of Nitrogen, i.e. 0.52 a.u., 1.61 a.u. and 3.92 a.u. with $\epsilon=\sqrt{1.02}$ and  $q_{ee}=\sqrt{0.5}$.   We will refer, therefore, to that model as Nitrogen.

\noindent{\it The Data Representation:}
The time evolution of TDSE\cite{suppl}  leads to events in which the three electrons are ionized. Can we extract a useful information on the dynamics from momenta distribution in such a case? We are interested in the triple ionization events, i.e. those events where, once the laser pulse have ceased, all three electrons are in the "outer" region. The 3D wavefunction obtained after the integration of the TDSE  allows us to learn about the momentum of outgoing electrons employing the technique of Dalitz plots.
These 
representation is instrumental in particle physics and have been proved to be very useful to disentangle electron-electron and electron-ion correlations in ion atom collisions
 \cite{Ciappina06,Schulz00}.  In its essence, the 3D wavefunction is projected onto the surrounding sphere by integrating over the radial coordinate, and then each octant of this sphere is projected onto a plane, forming an equilateral triangle. Each projection plane is chosen perpendicular to the diagonal line belonging to the particular octant; the type of projection is gnomonic with a tangent equal to the radius of the sphere. The borders between octants are formed by the $XY$, $XZ$ and $YZ$ plains, thus the vertices of the Dalitz plots correspond to exclusive motion of the $X$, $Y$ or $Z$ electron with the other two electrons at rest. We refer to these vertices as $X$, $Y$ and $Z$, respectively. For each Dalitz plot, an internal position is defined by a set of distances $\{ \pi_X,\pi_Y,\pi_Z \}$ to the sides of the triangle opposite to the $X,Y,Z$ vertices, correspondingly (see Fig. \ref{fig1} for illustration):

\begin{equation}
\pi_i = \left| \cfrac{p_i}{\sqrt{p_X^2+p_Y^2+p_Z^2}} \right|, \quad i=X,Y,Z,
\end{equation}

\noindent with $p_X,p_Y$ and $p_Z$ being the point coordinates in the momentum space. The particular octant, in turn, is defined by a set of quantities $(\xi_X,\xi_Y,\xi_Z)$:

\begin{equation}
\xi_i = \text{sign} \left( \cfrac{p_i}{\sqrt{p_X^2+p_Y^2+p_Z^2}} \right), \quad i=X,Y,Z,
\end{equation}

\noindent for the purpose of presentation we substitute $-1$  by $-$ and $+1$ by $+$. One needs to notice that by their very definition, Dalitz plots map ratios of the ejected electrons' momenta to the triple-ion momentum value, while information about momenta absolute values is lost. Still, however, one is able to determine the correlated escape of electrons, e.g. let us consider the plot representing the ($+++$) octant, then two electrons escaping with similar momenta, in the same direction along polarization axis, should contribute to the maxima that are equally distant to two chosen sides of the triangle, i.e. the maxima should appear along each of altitudes forming a structure with trifold symmetry due to electron indistinguishability.

\begin{figure}[t]
	\includegraphics[width=1.0\linewidth]{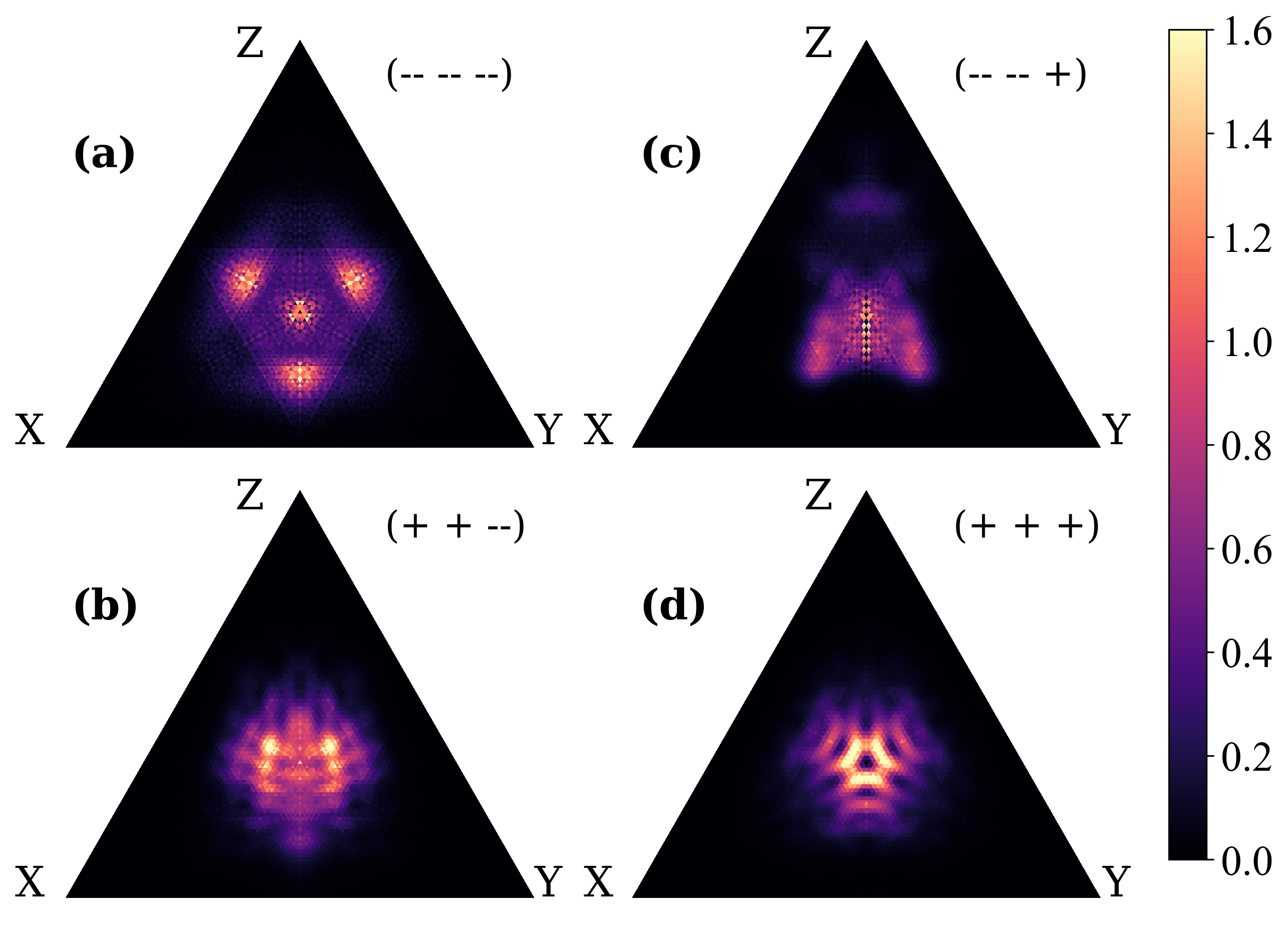}
	\caption{ (Color online) Experimental-type (spin direction averaged) Dalitz plots for triply-ionized state of Neon depict the relative momentum distribution of electrons released in different spatial directions with respect to the external field vector (``$+$'' for positive, ``$-$'' for negative). In the notation $(***)$ the first, second and third signs correspond to $X,Y$ and $Z$ electrons. The interpretation of the position in the plots is presented in Fig. \ref{fig1}. }
	\label{fig3}
\end{figure}

Let us now see the connection between the simulated data and the data of a prospective experiment. While an all-angles experiment allowing the detection of all ejected electrons independently of their direction in 3D space is a standard tool these days, the restricted-dimensionality code we use is not capable of resolving such an information. The numerical data do  allow one to determine the positive or negative direction of electronic motion with respect to the laser polarization axis only. On the other hand, in our calculation, we are able to extract spin-resolved information - a knowledge inaccessible in real experiments because the 
strong field phenomena in the non-relativistic regime are insensitive to the electron spin. Thus, in order to have a compatible set of data, we provide a Dalitz plots transformation to wash out any spin-resolved fingerprint. Such spin-averaged plots would correspond to angular-integrated (with respect to forth- "$+$" and back- "$-$" propagation) momentum-resolved 3-electron coincidence experiment.

The corresponding transformation that averages over spin directions is simple: for each experimentally-resolved combination of electrons [($+++$), ($++-$), ($--+$) and ($---$)] one collects all three possible orientation of electrons (vertices $XYZ, YZX, ZXY$) and then sums them up.
As a result, one is left with four plots per one simulation, compare Fig.~\ref{fig3}. Such a procedure is necessary for a system consisting of different spin directions. On the contrary, in the $p^3$ case no averaging is necessary \cite{suppl}.

\noindent{\it Results:}
In order to show the capabilities of an anticipated coincidence experiment on triple strong-field ionization, we simulated the dynamics within three-active-electron models of Nitrogen and Neon atoms -- the examples of systems with spatially fully antisymmetric wavefunction and with partially antisymmetric wavefunction, correspondingly. These two models were previously studied in the context of the ionization yield dependence on the laser field amplitude \cite{Thiede18,Efimov19,Efimov20,Prauzner-Bechcicki21}. In essence the results are as follows. At intermediate laser intensities, when the characteristic knee in the ionization yield is observed, there exist several channels leading to a triple ion, i.e. a sequential ionization, when each electron is liberated independently of others; a direct ionization, when all three electrons are set free at once; and a mixed ionization, when two electrons escape from the atom simultaneously and the third one is ejected separately (either single ionization follows double or vice versa). For atoms with fully antisymmetric wavefunction in their ground state the channels involving any kind of simultaneous escape (either the double ionization in the mixed triple ionization or the direct triple ionization) are suppressed in comparison to the sequential ionization channel. For atoms with a partially antisymmetric wavefunction, however, the mixed ionization channel is not suppressed \cite{Prauzner-Bechcicki21}. Basing on the fact that in the case of two-electron events the non-sequential ionization manifest its correlated character on the electron's momentum distributions in a form of the famous finger-like structures~\cite{staudte2007binary,rudenko2007correlated}, we expect to recognize indications of correlated motion in Dalitz plots as well. Furthermore, as the significance of diverse ionization channels in the case of Nitrogen and Neon is different, the correlated motion traces in Dalitz plots for Nitrogen and Neon, respectively, are expected to differ.

\begin{figure}[t]
	\includegraphics[width=1.0\linewidth]{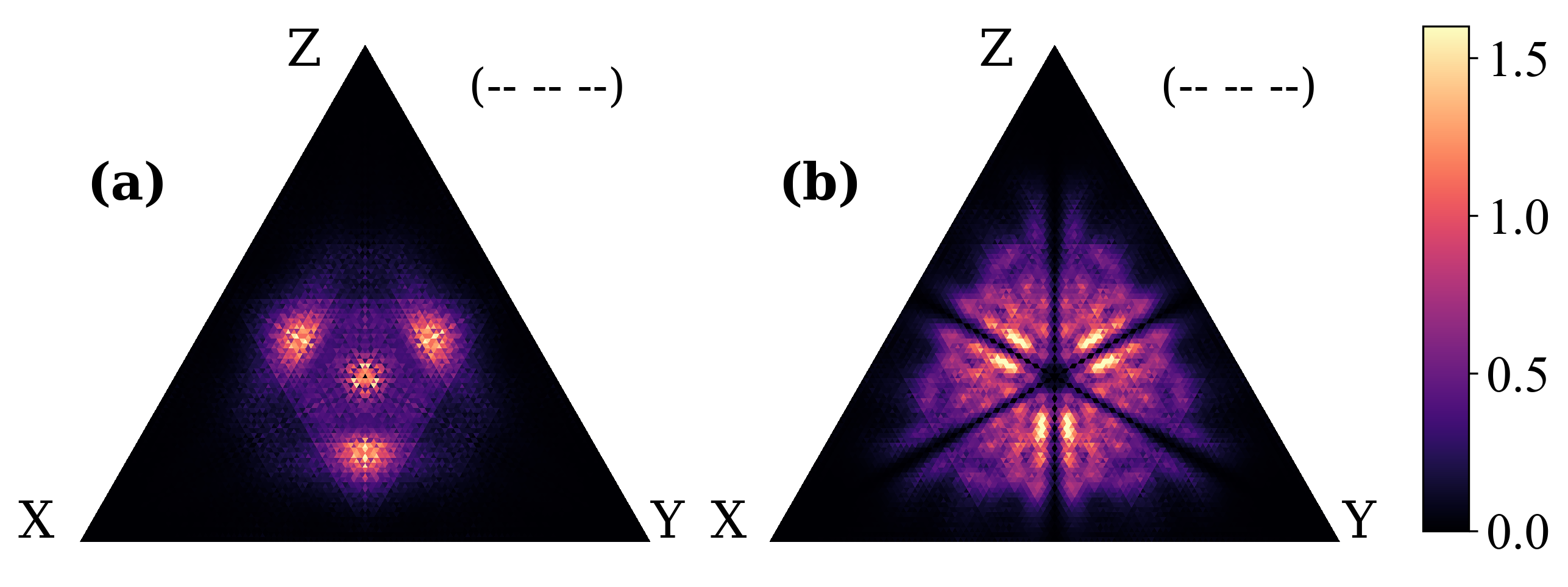}
	\caption{ (Color online) Dalitz plots for Neon (a) do not exhibit ``empty bisectors'' -- in contrast to those of Nitrogen (b). The plots are spin-averaged.  }
	\label{fignew1}
\end{figure}

\begin{figure}[t]
	\includegraphics[width=1.0\linewidth]{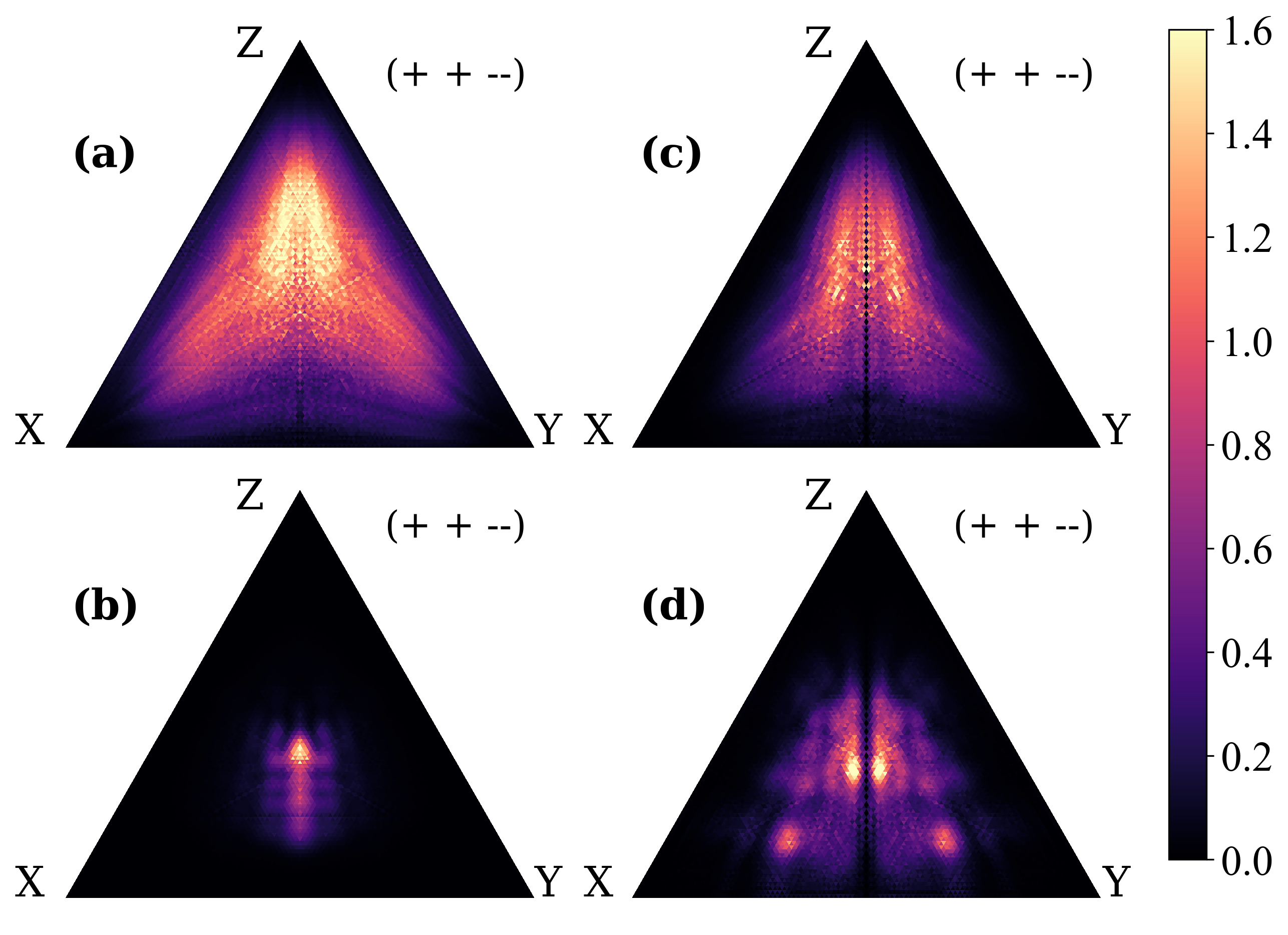}
	\caption{ (Color online) Dalitz plots for both Neon (a,b) and Nitrogen (c,d) experience qualitative change when moving from lower field values (a,c) to higher field values (b,d) around $F=0.1$ a.u. The plots are spin-averaged.}
	\label{fignew2}
\end{figure}

Fig.~\ref{fig3} shows a set of four Daliz plots obtained for Neon exposed to a pulse with amplitude $F=0.12$ a.u. One recognizes in panels (a) and (d), which collect momenta of electrons escaping in the same direction of the polarization axis, a structure centered at the triangle centroid that exhibits a tri-fold symmetry. The structure from panel (a) has three equivalent maxima corresponding to the motion of two electrons 
 with approximately equal momenta and the third one with significantly lower momentum - these maxima point to a mixed ionization: a direct double ionization accompanied by a single ionization event. 
 There is also a maximum right at the triangle centroid that corresponds to the direct triple escape. One could expect a similar structure in panel (d), as this panel collects momenta of electrons escaping in 
 opposite directions. It is not the case, because we limit ourselves to few-cycle optical cycles. On the contrary, for longer pulses, this should be observed. Panels (b) and (c) collect events in which two electrons 
 are moving in the same direction and the third one in the opposite one - 
 again the correlated motion of the former two electrons manifests itself in a structure that is aligned with and symmetric about the bisector falling from the vertex corresponding to the third electron moving in the opposite direction.

Let us now compare the Dalitz plots for Neon and Nitrogen. We limit our discussion to the octants representing electrons moving in the same direction with respect to the polarization axis.
Such plots are shown in Fig.~\ref{fignew1}. The difference in the observed structures is evident. Although in both panels the observed structures have tri-fold symmetry, in the case of Neon 
one observes clear maxima on three bisectors and at the triangle centroid (see panel (a)), in striking contrast to the zeros in the distributions along each of the bisectors that are visible for Nitrogen (see panel (b)). The spin symmetry clearly manifests itself in Dalitz plots. For Nitrogen, a spatial totally antisymmetric wavefunction has $\Psi = 0$ for $r_i=r_j, \, i,j=\{1,2,3\}$; this property is conserved during the transformation to the momentum space, leaving the diagonals empty. Such diagonals map to bisectors during the Dalitz plot creation. 
In the case of Neon, however, there is a {single} ``zero'' bisector in the Dalitz plots of the 8-plot set, thus it does not survive rotations and additions provided to obtain the ``experimental'' spin-averaged plot. The observed difference in Dalitz plots for Neon and Nitrogen is in line with recalled suppression of correlated escapes for the case of Nitrogen~\cite{Prauzner-Bechcicki21}.
At the same time both the N and Ne plots share a common feature: a qualitative change of Dalitz plot structures around the field amplitude $F=0.1$ a.u. (see Fig.~\ref{fignew2}) - distributions become narrower, centered at, aligned with and symmetric about the bisector falling from vertex corresponding to the electron moving in the opposite direction to the motion of the other two electrons. Comparison with ionization yields curves \cite{Thiede18,Prauzner-Bechcicki21}, directly implies a clear correlation between such a change and entering into the ``knee'' regime, where nonsequential ionization processes are instrumental.

\begin{figure}[t]
	\includegraphics[width=1.0\linewidth]{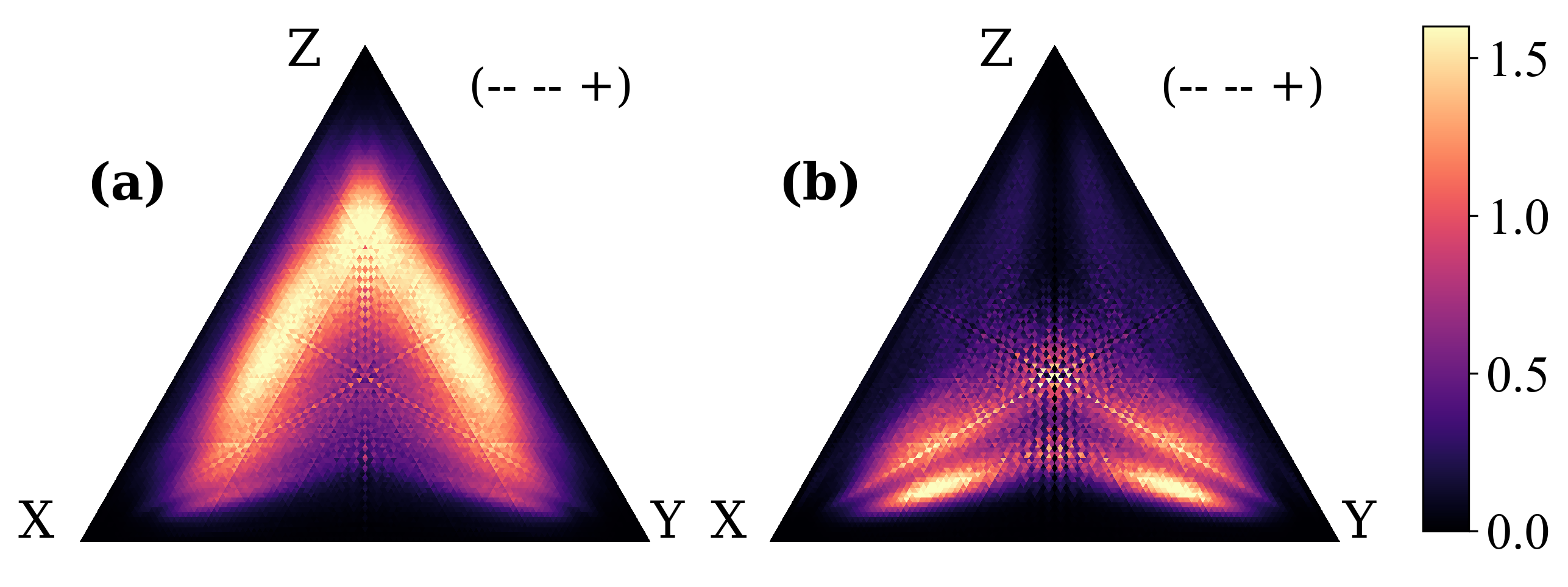}
	\caption{ (Color online) For the lower field values ($F<0.1$ a.u.) the $(--+)$ plots for Neon (a) and Nitrogen (b) show different shapes: the main structures are concentrated below the center of a Dalitz plot for the former case, and above this center in the latter case. The plots are spin-averaged. }
	\label{fignew3}
\end{figure}

There is one more remarkable difference between the Dalitz plots for N and Ne. In the intensity region below the knee, that is, about $F<0.1$ in our case, a noticable difference is seen in the general shapes of the structures in $(--+)$ octants (see Fig.~\ref{fignew3}). The maxima for N are all concentrated below the triangle centroid $O$ -- within the triangle $XOY$; the maxima for Ne, on the contrary, are located in the upper part of the plot, within triangles $ZOX$ and $ZOY$. The effect is robust and repeats for a number of intensities. Importantly, the absence of similar differences in $(++-)$ octants (see Fig.~\ref{fignew2} panels (a) and (c)) suggests a significant dependence of the signal on the carrier envelope phase (CEP), as should be expected for a few cycle pulse - study of such dependencies are in progress. Indeed, for the low field amplitude triple ionization would almost necessarily imply at least two cycles of ionization with subsequent recollision of an electron or two electrons. Efficiency of these sub-processes is inherently CEP-dependent. From the present data, the two same-direction electrons appear to be fastest in the case of N, while the opposite directions  electrons are most probable to have largest velocity for Ne.

\noindent{\it Summary:} The three-electron coincidence scheme appears to be an important tool for studying multiple electron ionization dynamics. We have proposed a possible experimental scheme for detecting electronic momenta after triple ionization of atoms by strong femtosecond laser pulses and have shown how the obtained data can be processed, visualized and analyzed. We have demonstrated how ionization of atoms with different spin symmetry of electronic valence shell can be visualized experimentally. The work in progress will analyze in detail the dependence of the Dalitz plots on the carrier-envelope phase which would provide a more detailed understanding of the process. Our simulations indicate a need for a theoretical model allowing one for a detailed interpretation of the Dalitz plots for three electron strong-field ionization.

\noindent{\it Acknowledgments} This work was initiated in discussions with late Bruno Eckhardt. We acknowledge discussions with Anne L'Huillier on the experimental feasibility of three-electron coincidence experiments. A support by  PL-Grid Infrastructure was vital for numerical results presented in this work.
It was realized under   National Science Centre (Poland) project Symfonia  No. 2016/20/W/ST4/00314.  M.L. acknowledges also 
support from ERC AdG NOQIA, Agencia Estatal de Investigación (“Severo Ochoa” Center of Excellence CEX2019-000910-S, Plan National FIDEUA PID2019-106901GB-I00/10.13039 / 501100011033, FPI), Fundació Privada Cellex, Fundació Mir-Puig, and from Generalitat de Catalunya (AGAUR Grant No. 2017 SGR 1341, CERCA program, QuantumCAT \_U16-011424 , co-funded by ERDF Operational Program of Catalonia 2014-2020), MINECO-EU QUANTERA MAQS (funded by State Research Agency (AEI) PCI2019-111828-2 / 10.13039/501100011033), EU Horizon 2020 FET-OPEN OPTOLogic (Grant No 899794), and the Marie Sklodowska-Curie grant STRETCH No 101029393.

\input{3emom.bbl}
\bigskip 
\section{Supplementary information to Three-electron correlations in strong laser field ionization: Spin induced effects}

We provide here more technical details on extracting momenta distributions from the wavefunctions obtained from the TDSE solution on a spacial grid. An extended description of the algorithm and its implementation will be published elsewhere. We show also exemplary Dalitz  plots obtained
 directly from the wavefunction analyzed, before the spin averaging procedure described in the main text.

\subsection{Extracting momenta distributions}	

As mentioned in the main text the ground states within an appropriate symmetry classes (different for Neon and Nitrogen) are obtained by an imaginary time propagation technique.  Ground states were computed on a 512 points grid for up to 100 a.u. of length in each direction. While this region seems quite large, it is necessary to correctly represent the tails of each ground state. At each time step, during the imaginary time propagation, the wavefunction was symmetrized back following the rules for a given symmetry class. In this way, the numerical inaccuracies would not drive us, e.g.. to lower lying symmetric (nonphysical) states. For real time propagation the grid has been expanded to 1024 points in each direction. 
  
Typically, in a real time propagation one introduces some complex absorbing potential close to the edges of the grid to avoid spurious reflection. The loss of the norm may then serve as an indicator of the ionization yield. Since we aim at momenta distributions, such a complex potential could artificially affect the electron momenta so we need to use a different approach. One possible way of approaching the problem would be to extend the grid sufficiently far 
so that the wavefunction never reaches the boundaries during the real time evolution. This is not feasible while keeping a reasonable grid step size. An alternative strategy has been devised in a seminal paper by Lein, Gross and Engel \cite{Lein00} for the two-electron case. We have also followed this scheme in our earlier studies \cite{Prauzner08}. The present algorithm is a simple (albeit technically hard) extension of this approach to three dimensions (electrons). The idea is as follows, all the space is divided into regions corresponding nominally to the Neutral atom (N), the Single (S), the Double (D) and the Triple (T) ions. During the time evolution one keeps track of the probability fluxes between different regions and that allows one to perform ionization dynamics simulations without absorbing boundaries at all. 

The basic physical idea employed in the method of Ref.~\cite{Lein00} is that an electron located far away from the parent ion, far beyond the classical point of no return $r_{t}=F/\omega^2$, would never again propagates close to the nucleus, thus will never again experience an influence of the ionic core as well as the companion electrons. Because of that its evolution is governed, to a good precision, exclusively by the kinetic energy operator $p_i^2/2$, where:
\begin{equation}
p_i^2/2 = \tilde{p}_i^2/2+(\sqrt{2/3})A(t)\tilde{p}_i,
\end{equation}
 where  $p_i$ is kinetical momentum, $\tilde{p}_i$ is canonical momentum and $A(t)$ is vector potential: $F(t)=-\partial A(t)/\partial t$. This way we neglect the influence of the nucleus as well as the brother electrons. Solution of the time-dependent Schr\"odinger equation in the dimension corresponding to this electron is analytic in the momentum representation; the wavefunction can be stored in momentum representation forever.  

\begin{figure}
	\includegraphics[width=1.0\linewidth]{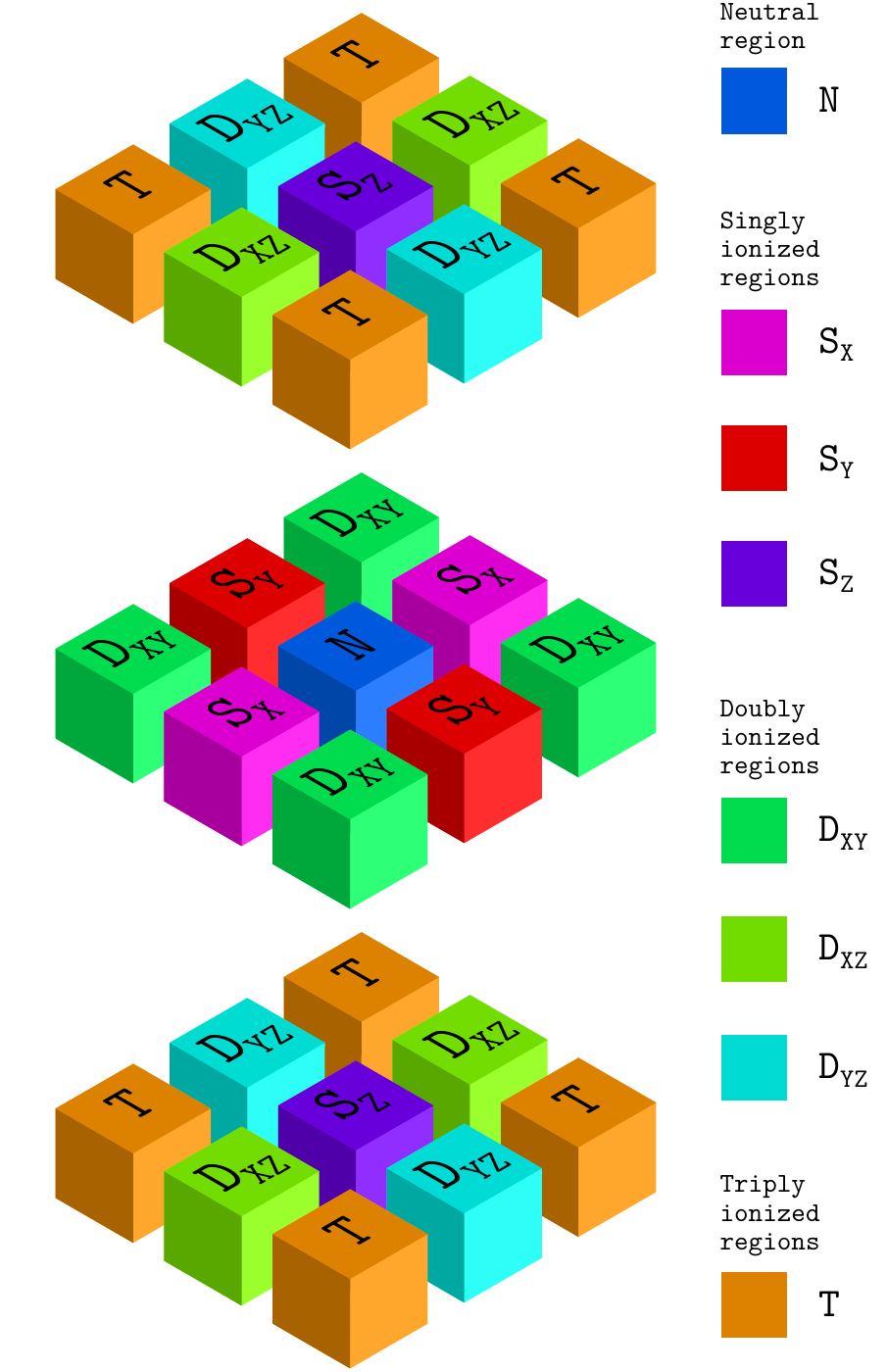}
	\caption{ (Color online) Space division. Cubes with the same name are incorporated into the same region. The position of cubes depicts relative position of the regions. The wavefunction transfer occurs between the neighboring regions. As far as none of the regions besides (N) allow storage of the wavefunction in position representation along all three dimensions, the cubes should not be interpreted as a space areas in the position representation. The boundary of the regions are contiguous to each other and are drawn detached in the figure for clarity purpose only.  }
	\label{fig1}
\end{figure}
\begin{figure}
	\includegraphics[width=1.0\linewidth]{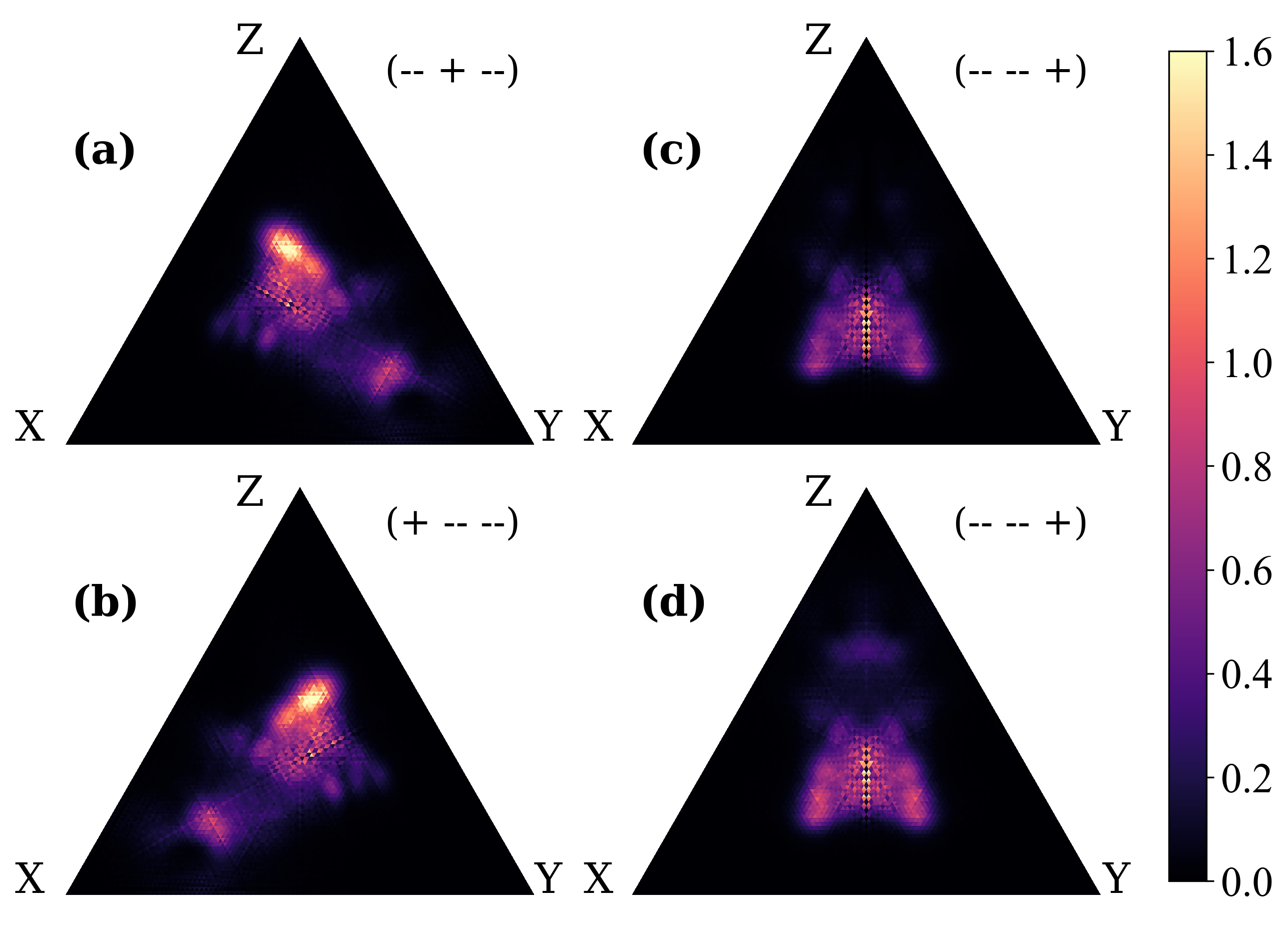}
	\caption{ (Color online) Spin sensitive plots (distinguishing the direction to which each spin is ejected --  (a,b,c)) are obtained by straightforward projection of the wavefunction; $X,Y,Z$ electrons are attributed with a particular spin. They are further concatenated to (d) after a preliminary rotation applied to each of them: we check that ``+'' becomes in position $Z$, so (a) is rotated counter clockwise, (b) - clockwise and (c) does not change its orientation. 
	}
	\label{fign}
\end{figure}

Following this idea we define a neutral atom region (N) as $\forall i:\, |r_i|<r_{t}$; single ion region ($S_k$) with $k$-th electron away from nucleus: $\forall i\ne k:\, |r_i|<r_{t}$, $|r_k|\geq r_{t}$; double ion region ($D_{kl}$) with $k$-th and $l$-th electrons away from nucleus: $\forall i\ne k,l:\, |r_i|<r_{t}$, $|r_k|\geq r_{t}$, $|r_l|\geq r_{t}$; triple ion region ($T$) with all electrons away from nucleus: $\forall i:\, |r_i|\geq r_{r}$. Graphical representation of the space division is presented in Fig.~\ref{fig1}.

Since we partially neglect interactions, a different effective Hamiltonian is solved in each region.  Hamiltonians are obtained by dropping the potential terms. They appear to be small enough due to the large values of the distances, taken as a parameter:
\begin{eqnarray}
&&H_N=\sum_{i=1}^3\frac{p_i^2}{2}+\sum_{i=1}^3V_{n-e}(r_i)+\sum_{i,j=1;i<j}^3V_{e-e}(r_i,r_j), \\
&&H_{S_k}=\sum_{i=1}^3\frac{p_i^2}{2}+\sum_{i=1,i\ne k}^3V_{n-e}(r_i)+V_{e-e}(r_i,r_j)|_{i,j\ne k}, \\
&&H_{D_{kl}}=\sum_{i=1}^3\frac{p_i^2}{2}+\sum_{i=1,i\ne k,l}^3V_{n-e}(r_i), \\
&&H_T=\sum_{i=1}^3\frac{p_i^2}{2}. 
\label{ph}
\end{eqnarray}
Note that the interaction with the laser field is hidden in the $p_i$'s. Therefore, in the implementation of the model, we use the velocity gauge Hamiltonian.

The Hamiltonians contain the kinetic and (restricted in some regions) potential terms. For calculating the wavefunction evolution during an incremental time step $\Delta t$, the split operator technique is used as in~\cite{Thiede18}, implying that the total solution of the TDSE is expressed as consequent solutions of the TDSE for the kinetic and potential operators. As solutions for these operators are analytical in momentum and position representations, respectively, the Fourier transform is applied to switch between them. In respective representations the actions of the kinetic and potential terms appear as a simple phase factor. If for a given region the $i$-coordinate the potential operators are absent, the $i$-dimension of the wavefunction is stored in the momentum representation exclusively. Thus, in the T region the wavefunction is $\psi(p_X,p_Y,p_Z)$ and directly yields the dominant ingredient of the three-dimensional momentum distribution as $|\psi(p_X,p_Y,p_Z)|^2$.

The continuity between different regions and the natural flow of the wavefunction through their boundaries is provided by a set of wavefunction-transfer operators, that are applied at each time step. In essence, during the action of such an operator, in each region, starting from (N) to (D)-s and excluding (T), the wavefunction is profiled. In this way, its parts close to the boundaries with the neighboring regions are separated and, after the corresponding partial Fourier transform, are coherently added to the wavefunctions of the neighboring regions. 

After ending the time evolution of the wavefunction, i.e.~when the external laser field ceased, a mask is applied to all the regions incorporated in the calculations.  The mask spatially extracts the ionized part of the wavefunction and transfers it to the neighboring regions: from N to S-regions, from S-regions to D-regions and from D-regions to the T one. Technically, masking is identical to the wave-transfer described previously with the only difference being the shape of the profile: the mask utilizes the boundary between areas corresponding to predominantly bound and predominantly free states. After application of the mask, one ends up with one N region, three S regions, three D regions and eight T regions corresponding to physically neutral, single-ionized, double-ionized and triple-ionized states. Thus, each region can be used for plotting momentum distributions independently: distributions for single ions from S regions, for double ions from D regions and for triple ions from T region with the corresponding one-, two- and three-electron distributions.

\subsection{Dalitz plots}

The construction of the Dalitz plots is described in the main text. As pointed out there, the experiments cannot yield spin-sensitive data,
thus the data from simulations have to be spin averaged as shown and described in the main text.
 Here we present an example of spin-sensitive numerical simulation - compare Fig. \ref{fign}.

The resulting plots imply that the unique non-symmetric pattern of Dalitz plots for Neon (partially antisymmetric spin configuration) cannot be accessed with the experimental data that provide a two-fold symmetry for ($+--$) and ($++-$) data for both Neon and Nitrogen -- see example in Fig. \ref{fign}.

\end{document}

%% file: 3emom.bbl
\providecommand{\noopsort}[1]{}\providecommand{\singleletter}[1]{#1}%

%% file: 3emom.bbl
\begin{thebibliography}{63}%
\makeatletter
\providecommand \@ifxundefined [1]{%
 \@ifx{#1\undefined}
}%
\providecommand \@ifnum [1]{%
 \ifnum #1\expandafter \@firstoftwo
 \else \expandafter \@secondoftwo
 \fi
}%
\providecommand \@ifx [1]{%
 \ifx #1\expandafter \@firstoftwo
 \else \expandafter \@secondoftwo
 \fi
}%
\providecommand \natexlab [1]{#1}%
\providecommand \enquote  [1]{``#1''}%
\providecommand \bibnamefont  [1]{#1}%
\providecommand \bibfnamefont [1]{#1}%
\providecommand \citenamefont [1]{#1}%
\providecommand \href@noop [0]{\@secondoftwo}%
\providecommand \href [0]{\begingroup \@sanitize@url \@href}%
\providecommand \@href[1]{\@@startlink{#1}\@@href}%
\providecommand \@@href[1]{\endgroup#1\@@endlink}%
\providecommand \@sanitize@url [0]{\catcode `\\12\catcode `\$12\catcode
  `\&12\catcode `\#12\catcode `\^12\catcode `\_12\catcode `\%12\relax}%
\providecommand \@@startlink[1]{}%
\providecommand \@@endlink[0]{}%
\providecommand \url  [0]{\begingroup\@sanitize@url \@url }%
\providecommand \@url [1]{\endgroup\@href {#1}{\urlprefix }}%
\providecommand \urlprefix  [0]{URL }%
\providecommand \Eprint [0]{\href }%
\providecommand \doibase [0]{http://dx.doi.org/}%
\providecommand \selectlanguage [0]{\@gobble}%
\providecommand \bibinfo  [0]{\@secondoftwo}%
\providecommand \bibfield  [0]{\@secondoftwo}%
\providecommand \translation [1]{[#1]}%
\providecommand \BibitemOpen [0]{}%
\providecommand \bibitemStop [0]{}%
\providecommand \bibitemNoStop [0]{.\EOS\space}%
\providecommand \EOS [0]{\spacefactor3000\relax}%
\providecommand \BibitemShut  [1]{\csname bibitem#1\endcsname}%
\let\auto@bib@innerbib\@empty
\bibitem [{\citenamefont {Krausz}\ and\ \citenamefont
  {Ivanov}(2009)}]{Krausz09}%
  \BibitemOpen
  \bibfield  {author} {\bibinfo {author} {\bibfnamefont {F.}~\bibnamefont
  {Krausz}}\ and\ \bibinfo {author} {\bibfnamefont {M.}~\bibnamefont
  {Ivanov}},\ }\href {\doibase 10.1103/RevModPhys.81.163} {\bibfield  {journal}
  {\bibinfo  {journal} {Rev. Mod. Phys.}\ }\textbf {\bibinfo {volume} {81}},\
  \bibinfo {pages} {163} (\bibinfo {year} {2009})}\BibitemShut {NoStop}%
\bibitem [{\citenamefont {L'Huillier}\ \emph {et~al.}(1982)\citenamefont
  {L'Huillier}, \citenamefont {Lompre}, \citenamefont {Mainfray},\ and\
  \citenamefont {Manus}}]{Lhuillier82}%
  \BibitemOpen
  \bibfield  {author} {\bibinfo {author} {\bibfnamefont {A.}~\bibnamefont
  {L'Huillier}}, \bibinfo {author} {\bibfnamefont {L.~A.}\ \bibnamefont
  {Lompre}}, \bibinfo {author} {\bibfnamefont {G.}~\bibnamefont {Mainfray}}, \
  and\ \bibinfo {author} {\bibfnamefont {C.}~\bibnamefont {Manus}},\ }\href
  {\doibase 10.1103/PhysRevLett.48.1814} {\bibfield  {journal} {\bibinfo
  {journal} {Phys. Rev. Lett.}\ }\textbf {\bibinfo {volume} {48}},\ \bibinfo
  {pages} {1814} (\bibinfo {year} {1982})}\BibitemShut {NoStop}%
\bibitem [{\citenamefont {l'Huillier}\ \emph {et~al.}(1983)\citenamefont
  {l'Huillier}, \citenamefont {Lompre}, \citenamefont {Mainfray},\ and\
  \citenamefont {Manus}}]{Lhuillier83}%
  \BibitemOpen
  \bibfield  {author} {\bibinfo {author} {\bibfnamefont {A.}~\bibnamefont
  {l'Huillier}}, \bibinfo {author} {\bibfnamefont {L.~A.}\ \bibnamefont
  {Lompre}}, \bibinfo {author} {\bibfnamefont {G.}~\bibnamefont {Mainfray}}, \
  and\ \bibinfo {author} {\bibfnamefont {C.}~\bibnamefont {Manus}},\ }\href
  {\doibase 10.1103/PhysRevA.27.2503} {\bibfield  {journal} {\bibinfo
  {journal} {Phys. Rev. A}\ }\textbf {\bibinfo {volume} {27}},\ \bibinfo
  {pages} {2503} (\bibinfo {year} {1983})}\BibitemShut {NoStop}%
\bibitem [{\citenamefont {Luk}\ \emph {et~al.}(1983)\citenamefont {Luk},
  \citenamefont {Pummer}, \citenamefont {Boyer}, \citenamefont {Shahidi},
  \citenamefont {Egger},\ and\ \citenamefont {Rhodes}}]{Luk83}%
  \BibitemOpen
  \bibfield  {author} {\bibinfo {author} {\bibfnamefont {T.~S.}\ \bibnamefont
  {Luk}}, \bibinfo {author} {\bibfnamefont {H.}~\bibnamefont {Pummer}},
  \bibinfo {author} {\bibfnamefont {K.}~\bibnamefont {Boyer}}, \bibinfo
  {author} {\bibfnamefont {M.}~\bibnamefont {Shahidi}}, \bibinfo {author}
  {\bibfnamefont {H.}~\bibnamefont {Egger}}, \ and\ \bibinfo {author}
  {\bibfnamefont {C.~K.}\ \bibnamefont {Rhodes}},\ }\href {\doibase
  10.1103/PhysRevLett.51.110} {\bibfield  {journal} {\bibinfo  {journal} {Phys.
  Rev. Lett.}\ }\textbf {\bibinfo {volume} {51}},\ \bibinfo {pages} {110}
  (\bibinfo {year} {1983})}\BibitemShut {NoStop}%
\bibitem [{\citenamefont {Boyer}\ \emph {et~al.}(1984)\citenamefont {Boyer},
  \citenamefont {Egger}, \citenamefont {Luk}, \citenamefont {Pummer},\ and\
  \citenamefont {Rhodes}}]{Boyer84}%
  \BibitemOpen
  \bibfield  {author} {\bibinfo {author} {\bibfnamefont {K.}~\bibnamefont
  {Boyer}}, \bibinfo {author} {\bibfnamefont {H.}~\bibnamefont {Egger}},
  \bibinfo {author} {\bibfnamefont {T.~S.}\ \bibnamefont {Luk}}, \bibinfo
  {author} {\bibfnamefont {H.}~\bibnamefont {Pummer}}, \ and\ \bibinfo {author}
  {\bibfnamefont {C.~K.}\ \bibnamefont {Rhodes}},\ }\href {\doibase
  10.1364/JOSAB.1.000003} {\bibfield  {journal} {\bibinfo  {journal} {J. Opt.
  Soc. Am. B}\ }\textbf {\bibinfo {volume} {1}},\ \bibinfo {pages} {3}
  (\bibinfo {year} {1984})}\BibitemShut {NoStop}%
\bibitem [{\citenamefont {Luk}\ \emph {et~al.}(1985)\citenamefont {Luk},
  \citenamefont {Johann}, \citenamefont {Egger}, \citenamefont {Pummer},\ and\
  \citenamefont {Rhodes}}]{Luk85}%
  \BibitemOpen
  \bibfield  {author} {\bibinfo {author} {\bibfnamefont {T.~S.}\ \bibnamefont
  {Luk}}, \bibinfo {author} {\bibfnamefont {U.}~\bibnamefont {Johann}},
  \bibinfo {author} {\bibfnamefont {H.}~\bibnamefont {Egger}}, \bibinfo
  {author} {\bibfnamefont {H.}~\bibnamefont {Pummer}}, \ and\ \bibinfo {author}
  {\bibfnamefont {C.~K.}\ \bibnamefont {Rhodes}},\ }\href {\doibase
  10.1103/PhysRevA.32.214} {\bibfield  {journal} {\bibinfo  {journal} {Phys.
  Rev. A}\ }\textbf {\bibinfo {volume} {32}},\ \bibinfo {pages} {214} (\bibinfo
  {year} {1985})}\BibitemShut {NoStop}%
\bibitem [{\citenamefont {Chin}\ \emph {et~al.}(1985)\citenamefont {Chin},
  \citenamefont {Yergeau},\ and\ \citenamefont {Lavigne}}]{Chin85}%
  \BibitemOpen
  \bibfield  {author} {\bibinfo {author} {\bibfnamefont {S.~L.}\ \bibnamefont
  {Chin}}, \bibinfo {author} {\bibfnamefont {F.}~\bibnamefont {Yergeau}}, \
  and\ \bibinfo {author} {\bibfnamefont {P.}~\bibnamefont {Lavigne}},\ }\href
  {\doibase 10.1088/0022-3700/18/8/001} {\bibfield  {journal} {\bibinfo
  {journal} {Journal of Physics B: Atomic and Molecular Physics}\ }\textbf
  {\bibinfo {volume} {18}},\ \bibinfo {pages} {L213} (\bibinfo {year}
  {1985})}\BibitemShut {NoStop}%
\bibitem [{\citenamefont {Lambropoulos}(1985)}]{Lambropoulos85}%
  \BibitemOpen
  \bibfield  {author} {\bibinfo {author} {\bibfnamefont {P.}~\bibnamefont
  {Lambropoulos}},\ }\href {\doibase 10.1103/PhysRevLett.55.2141} {\bibfield
  {journal} {\bibinfo  {journal} {Phys. Rev. Lett.}\ }\textbf {\bibinfo
  {volume} {55}},\ \bibinfo {pages} {2141} (\bibinfo {year}
  {1985})}\BibitemShut {NoStop}%
\bibitem [{\citenamefont {Yergeau}\ \emph {et~al.}(1987)\citenamefont
  {Yergeau}, \citenamefont {Chin},\ and\ \citenamefont {Lavigne}}]{Yergeau87}%
  \BibitemOpen
  \bibfield  {author} {\bibinfo {author} {\bibfnamefont {F.}~\bibnamefont
  {Yergeau}}, \bibinfo {author} {\bibfnamefont {S.~L.}\ \bibnamefont {Chin}}, \
  and\ \bibinfo {author} {\bibfnamefont {P.}~\bibnamefont {Lavigne}},\ }\href
  {\doibase 10.1088/0022-3700/20/4/013} {\bibfield  {journal} {\bibinfo
  {journal} {Journal of Physics B: Atomic and Molecular Physics}\ }\textbf
  {\bibinfo {volume} {20}},\ \bibinfo {pages} {723} (\bibinfo {year}
  {1987})}\BibitemShut {NoStop}%
\bibitem [{\citenamefont {Crance}(1987)}]{Crance87}%
  \BibitemOpen
  \bibfield  {author} {\bibinfo {author} {\bibfnamefont {M.}~\bibnamefont
  {Crance}},\ }\href {\doibase https://doi.org/10.1016/S0370-1573(87)80003-0}
  {\bibfield  {journal} {\bibinfo  {journal} {Physics Reports}\ }\textbf
  {\bibinfo {volume} {144}},\ \bibinfo {pages} {118} (\bibinfo {year}
  {1987})}\BibitemShut {NoStop}%
\bibitem [{\citenamefont {Mu}\ \emph {et~al.}(1986)\citenamefont {Mu},
  \citenamefont {\AA{}berg}, \citenamefont {Blomberg},\ and\ \citenamefont
  {Crasemann}}]{Mu86}%
  \BibitemOpen
  \bibfield  {author} {\bibinfo {author} {\bibfnamefont {X.-D.}\ \bibnamefont
  {Mu}}, \bibinfo {author} {\bibfnamefont {T.}~\bibnamefont {\AA{}berg}},
  \bibinfo {author} {\bibfnamefont {A.}~\bibnamefont {Blomberg}}, \ and\
  \bibinfo {author} {\bibfnamefont {B.}~\bibnamefont {Crasemann}},\ }\href
  {\doibase 10.1103/PhysRevLett.56.1909} {\bibfield  {journal} {\bibinfo
  {journal} {Phys. Rev. Lett.}\ }\textbf {\bibinfo {volume} {56}},\ \bibinfo
  {pages} {1909} (\bibinfo {year} {1986})}\BibitemShut {NoStop}%
\bibitem [{\citenamefont {\AA{}berg}\ \emph {et~al.}(1984)\citenamefont
  {\AA{}berg}, \citenamefont {Blomberg}, \citenamefont {Tulkki},\ and\
  \citenamefont {Goscinski}}]{Aberg84}%
  \BibitemOpen
  \bibfield  {author} {\bibinfo {author} {\bibfnamefont {T.}~\bibnamefont
  {\AA{}berg}}, \bibinfo {author} {\bibfnamefont {A.}~\bibnamefont {Blomberg}},
  \bibinfo {author} {\bibfnamefont {J.}~\bibnamefont {Tulkki}}, \ and\ \bibinfo
  {author} {\bibfnamefont {O.}~\bibnamefont {Goscinski}},\ }\href {\doibase
  10.1103/PhysRevLett.52.1207} {\bibfield  {journal} {\bibinfo  {journal}
  {Phys. Rev. Lett.}\ }\textbf {\bibinfo {volume} {52}},\ \bibinfo {pages}
  {1207} (\bibinfo {year} {1984})}\BibitemShut {NoStop}%
\bibitem [{\citenamefont {Geltman}(1985)}]{Geltman85}%
  \BibitemOpen
  \bibfield  {author} {\bibinfo {author} {\bibfnamefont {S.}~\bibnamefont
  {Geltman}},\ }\href {\doibase 10.1103/PhysRevLett.54.1909} {\bibfield
  {journal} {\bibinfo  {journal} {Phys. Rev. Lett.}\ }\textbf {\bibinfo
  {volume} {54}},\ \bibinfo {pages} {1909} (\bibinfo {year}
  {1985})}\BibitemShut {NoStop}%
\bibitem [{\citenamefont {Zakrzewski}(1986)}]{Zakrzewski86}%
  \BibitemOpen
  \bibfield  {author} {\bibinfo {author} {\bibfnamefont {J.}~\bibnamefont
  {Zakrzewski}},\ }\href {\doibase 10.1088/0022-3700/19/9/004} {\bibfield
  {journal} {\bibinfo  {journal} {Journal of Physics B: Atomic and Molecular
  Physics}\ }\textbf {\bibinfo {volume} {19}},\ \bibinfo {pages} {L315}
  (\bibinfo {year} {1986})}\BibitemShut {NoStop}%
\bibitem [{\citenamefont {Lewenstein}(1986)}]{Lewenstein86}%
  \BibitemOpen
  \bibfield  {author} {\bibinfo {author} {\bibfnamefont {M.}~\bibnamefont
  {Lewenstein}},\ }\href {\doibase 10.1088/0022-3700/19/9/003} {\bibfield
  {journal} {\bibinfo  {journal} {Journal of Physics B: Atomic and Molecular
  Physics}\ }\textbf {\bibinfo {volume} {19}},\ \bibinfo {pages} {L309}
  (\bibinfo {year} {1986})}\BibitemShut {NoStop}%
\bibitem [{\citenamefont {Geltman}\ and\ \citenamefont
  {Zakrzewski}(1988)}]{Geltman88}%
  \BibitemOpen
  \bibfield  {author} {\bibinfo {author} {\bibfnamefont {S.}~\bibnamefont
  {Geltman}}\ and\ \bibinfo {author} {\bibfnamefont {J.}~\bibnamefont
  {Zakrzewski}},\ }\href {\doibase 10.1088/0953-4075/21/1/005} {\bibfield
  {journal} {\bibinfo  {journal} {Journal of Physics B: Atomic, Molecular and
  Optical Physics}\ }\textbf {\bibinfo {volume} {21}},\ \bibinfo {pages} {47}
  (\bibinfo {year} {1988})}\BibitemShut {NoStop}%
\bibitem [{\citenamefont {Fittinghoff}\ \emph {et~al.}(1992)\citenamefont
  {Fittinghoff}, \citenamefont {Bolton}, \citenamefont {Chang},\ and\
  \citenamefont {Kulander}}]{Fittinghoff92}%
  \BibitemOpen
  \bibfield  {author} {\bibinfo {author} {\bibfnamefont {D.~N.}\ \bibnamefont
  {Fittinghoff}}, \bibinfo {author} {\bibfnamefont {P.~R.}\ \bibnamefont
  {Bolton}}, \bibinfo {author} {\bibfnamefont {B.}~\bibnamefont {Chang}}, \
  and\ \bibinfo {author} {\bibfnamefont {K.~C.}\ \bibnamefont {Kulander}},\
  }\href {\doibase 10.1103/PhysRevLett.69.2642} {\bibfield  {journal} {\bibinfo
   {journal} {Phys. Rev. Lett.}\ }\textbf {\bibinfo {volume} {69}},\ \bibinfo
  {pages} {2642} (\bibinfo {year} {1992})}\BibitemShut {NoStop}%
\bibitem [{\citenamefont {Kondo}\ \emph {et~al.}(1993)\citenamefont {Kondo},
  \citenamefont {Sagisaka}, \citenamefont {Tamida}, \citenamefont {Nabekawa},\
  and\ \citenamefont {Watanabe}}]{Kondo93}%
  \BibitemOpen
  \bibfield  {author} {\bibinfo {author} {\bibfnamefont {K.}~\bibnamefont
  {Kondo}}, \bibinfo {author} {\bibfnamefont {A.}~\bibnamefont {Sagisaka}},
  \bibinfo {author} {\bibfnamefont {T.}~\bibnamefont {Tamida}}, \bibinfo
  {author} {\bibfnamefont {Y.}~\bibnamefont {Nabekawa}}, \ and\ \bibinfo
  {author} {\bibfnamefont {S.}~\bibnamefont {Watanabe}},\ }\href {\doibase
  10.1103/PhysRevA.48.R2531} {\bibfield  {journal} {\bibinfo  {journal} {Phys.
  Rev. A}\ }\textbf {\bibinfo {volume} {48}},\ \bibinfo {pages} {R2531}
  (\bibinfo {year} {1993})}\BibitemShut {NoStop}%
\bibitem [{\citenamefont {Walker}\ \emph {et~al.}(1994)\citenamefont {Walker},
  \citenamefont {Sheehy}, \citenamefont {DiMauro}, \citenamefont {Agostini},
  \citenamefont {Schafer},\ and\ \citenamefont
  {Kulander}}]{walker1994precision}%
  \BibitemOpen
  \bibfield  {author} {\bibinfo {author} {\bibfnamefont {B.}~\bibnamefont
  {Walker}}, \bibinfo {author} {\bibfnamefont {B.}~\bibnamefont {Sheehy}},
  \bibinfo {author} {\bibfnamefont {L.~F.}\ \bibnamefont {DiMauro}}, \bibinfo
  {author} {\bibfnamefont {P.}~\bibnamefont {Agostini}}, \bibinfo {author}
  {\bibfnamefont {K.~J.}\ \bibnamefont {Schafer}}, \ and\ \bibinfo {author}
  {\bibfnamefont {K.~C.}\ \bibnamefont {Kulander}},\ }\href {\doibase
  10.1103/PhysRevLett.73.1227} {\bibfield  {journal} {\bibinfo  {journal}
  {Phys. Rev. Lett.}\ }\textbf {\bibinfo {volume} {73}},\ \bibinfo {pages}
  {1227} (\bibinfo {year} {1994})}\BibitemShut {NoStop}%
\bibitem [{\citenamefont {Corkum}(1993)}]{Corkum93}%
  \BibitemOpen
  \bibfield  {author} {\bibinfo {author} {\bibfnamefont {P.~B.}\ \bibnamefont
  {Corkum}},\ }\href {\doibase 10.1103/PhysRevLett.71.1994} {\bibfield
  {journal} {\bibinfo  {journal} {Phys. Rev. Lett.}\ }\textbf {\bibinfo
  {volume} {71}},\ \bibinfo {pages} {1994} (\bibinfo {year}
  {1993})}\BibitemShut {NoStop}%
\bibitem [{\citenamefont {Eberly}\ \emph {et~al.}(1991)\citenamefont {Eberly},
  \citenamefont {Javanainen},\ and\ \citenamefont
  {Rz\c{a}\.zewski}}]{Eberly91}%
  \BibitemOpen
  \bibfield  {author} {\bibinfo {author} {\bibfnamefont {J.}~\bibnamefont
  {Eberly}}, \bibinfo {author} {\bibfnamefont {J.}~\bibnamefont {Javanainen}},
  \ and\ \bibinfo {author} {\bibfnamefont {K.}~\bibnamefont
  {Rz\c{a}\.zewski}},\ }\href {\doibase
  https://doi.org/10.1016/0370-1573(91)90131-5} {\bibfield  {journal} {\bibinfo
   {journal} {Physics Reports}\ }\textbf {\bibinfo {volume} {204}},\ \bibinfo
  {pages} {331} (\bibinfo {year} {1991})}\BibitemShut {NoStop}%
\bibitem [{\citenamefont {Lewenstein}\ \emph {et~al.}(1994)\citenamefont
  {Lewenstein}, \citenamefont {Balcou}, \citenamefont {Ivanov}, \citenamefont
  {L'Huillier},\ and\ \citenamefont {Corkum}}]{Lewenstein94}%
  \BibitemOpen
  \bibfield  {author} {\bibinfo {author} {\bibfnamefont {M.}~\bibnamefont
  {Lewenstein}}, \bibinfo {author} {\bibfnamefont {P.}~\bibnamefont {Balcou}},
  \bibinfo {author} {\bibfnamefont {M.~Y.}\ \bibnamefont {Ivanov}}, \bibinfo
  {author} {\bibfnamefont {A.}~\bibnamefont {L'Huillier}}, \ and\ \bibinfo
  {author} {\bibfnamefont {P.~B.}\ \bibnamefont {Corkum}},\ }\href {\doibase
  10.1103/PhysRevA.49.2117} {\bibfield  {journal} {\bibinfo  {journal} {Phys.
  Rev. A}\ }\textbf {\bibinfo {volume} {49}},\ \bibinfo {pages} {2117}
  (\bibinfo {year} {1994})}\BibitemShut {NoStop}%
\bibitem [{\citenamefont {Dörner}\ \emph {et~al.}(2000)\citenamefont
  {Dörner}, \citenamefont {Mergel}, \citenamefont {Jagutzki}, \citenamefont
  {Spielberger}, \citenamefont {Ullrich}, \citenamefont {Moshammer},\ and\
  \citenamefont {Schmidt-Böcking}}]{dorner00}%
  \BibitemOpen
  \bibfield  {author} {\bibinfo {author} {\bibfnamefont {R.}~\bibnamefont
  {Dörner}}, \bibinfo {author} {\bibfnamefont {V.}~\bibnamefont {Mergel}},
  \bibinfo {author} {\bibfnamefont {O.}~\bibnamefont {Jagutzki}}, \bibinfo
  {author} {\bibfnamefont {L.}~\bibnamefont {Spielberger}}, \bibinfo {author}
  {\bibfnamefont {J.}~\bibnamefont {Ullrich}}, \bibinfo {author} {\bibfnamefont
  {R.}~\bibnamefont {Moshammer}}, \ and\ \bibinfo {author} {\bibfnamefont
  {H.}~\bibnamefont {Schmidt-Böcking}},\ }\href {\doibase
  https://doi.org/10.1016/S0370-1573(99)00109-X} {\bibfield  {journal}
  {\bibinfo  {journal} {Physics Reports}\ }\textbf {\bibinfo {volume} {330}},\
  \bibinfo {pages} {95} (\bibinfo {year} {2000})}\BibitemShut {NoStop}%
\bibitem [{\citenamefont {Weber}\ \emph {et~al.}(2000)\citenamefont {Weber},
  \citenamefont {Giessen}, \citenamefont {Weckenbrock}, \citenamefont
  {Urbasch}, \citenamefont {Staudte}, \citenamefont {Spielberger},
  \citenamefont {Jagutzki}, \citenamefont {Mergel}, \citenamefont {Vollmer},\
  and\ \citenamefont {D{\"o}rner}}]{weber00}%
  \BibitemOpen
  \bibfield  {author} {\bibinfo {author} {\bibfnamefont {T.}~\bibnamefont
  {Weber}}, \bibinfo {author} {\bibfnamefont {H.}~\bibnamefont {Giessen}},
  \bibinfo {author} {\bibfnamefont {M.}~\bibnamefont {Weckenbrock}}, \bibinfo
  {author} {\bibfnamefont {G.}~\bibnamefont {Urbasch}}, \bibinfo {author}
  {\bibfnamefont {A.}~\bibnamefont {Staudte}}, \bibinfo {author} {\bibfnamefont
  {L.}~\bibnamefont {Spielberger}}, \bibinfo {author} {\bibfnamefont
  {O.}~\bibnamefont {Jagutzki}}, \bibinfo {author} {\bibfnamefont
  {V.}~\bibnamefont {Mergel}}, \bibinfo {author} {\bibfnamefont
  {M.}~\bibnamefont {Vollmer}}, \ and\ \bibinfo {author} {\bibfnamefont
  {R.}~\bibnamefont {D{\"o}rner}},\ }\href@noop {} {\bibfield  {journal}
  {\bibinfo  {journal} {Nature}\ }\textbf {\bibinfo {volume} {405}},\ \bibinfo
  {pages} {658} (\bibinfo {year} {2000})}\BibitemShut {NoStop}%
\bibitem [{\citenamefont {Moshammer}\ \emph {et~al.}(2000)\citenamefont
  {Moshammer}, \citenamefont {Feuerstein}, \citenamefont {Schmitt},
  \citenamefont {Dorn}, \citenamefont {Schroter}, \citenamefont {Ullrich},
  \citenamefont {Rottke}, \citenamefont {Trump}, \citenamefont {Wittmann},
  \citenamefont {Korn}, \citenamefont {Hoffmann},\ and\ \citenamefont
  {Sandner}}]{Moshammer2000-vn}%
  \BibitemOpen
  \bibfield  {author} {\bibinfo {author} {\bibfnamefont {R.}~\bibnamefont
  {Moshammer}}, \bibinfo {author} {\bibfnamefont {B.}~\bibnamefont
  {Feuerstein}}, \bibinfo {author} {\bibfnamefont {W.}~\bibnamefont {Schmitt}},
  \bibinfo {author} {\bibfnamefont {A.}~\bibnamefont {Dorn}}, \bibinfo {author}
  {\bibfnamefont {C.~D.}\ \bibnamefont {Schroter}}, \bibinfo {author}
  {\bibfnamefont {J.}~\bibnamefont {Ullrich}}, \bibinfo {author} {\bibfnamefont
  {H.}~\bibnamefont {Rottke}}, \bibinfo {author} {\bibfnamefont
  {C.}~\bibnamefont {Trump}}, \bibinfo {author} {\bibfnamefont
  {M.}~\bibnamefont {Wittmann}}, \bibinfo {author} {\bibfnamefont
  {G.}~\bibnamefont {Korn}}, \bibinfo {author} {\bibfnamefont {K.}~\bibnamefont
  {Hoffmann}}, \ and\ \bibinfo {author} {\bibfnamefont {W.}~\bibnamefont
  {Sandner}},\ }\href {\doibase 10.1103/PhysRevLett.84.447} {\bibfield
  {journal} {\bibinfo  {journal} {Phys. Rev. Lett.}\ }\textbf {\bibinfo
  {volume} {84}},\ \bibinfo {pages} {447} (\bibinfo {year} {2000})}\BibitemShut
  {NoStop}%
\bibitem [{\citenamefont {Liu}\ \emph {et~al.}(2008)\citenamefont {Liu},
  \citenamefont {Tschuch}, \citenamefont {Rudenko}, \citenamefont {D{\"u}rr},
  \citenamefont {Siegel}, \citenamefont {Morgner}, \citenamefont {Moshammer},\
  and\ \citenamefont {Ullrich}}]{Liu2008-yt}%
  \BibitemOpen
  \bibfield  {author} {\bibinfo {author} {\bibfnamefont {Y.}~\bibnamefont
  {Liu}}, \bibinfo {author} {\bibfnamefont {S.}~\bibnamefont {Tschuch}},
  \bibinfo {author} {\bibfnamefont {A.}~\bibnamefont {Rudenko}}, \bibinfo
  {author} {\bibfnamefont {M.}~\bibnamefont {D{\"u}rr}}, \bibinfo {author}
  {\bibfnamefont {M.}~\bibnamefont {Siegel}}, \bibinfo {author} {\bibfnamefont
  {U.}~\bibnamefont {Morgner}}, \bibinfo {author} {\bibfnamefont
  {R.}~\bibnamefont {Moshammer}}, \ and\ \bibinfo {author} {\bibfnamefont
  {J.}~\bibnamefont {Ullrich}},\ }\href {\doibase
  10.1103/PhysRevLett.101.053001} {\bibfield  {journal} {\bibinfo  {journal}
  {Phys. Rev. Lett.}\ }\textbf {\bibinfo {volume} {101}},\ \bibinfo {pages}
  {053001} (\bibinfo {year} {2008})}\BibitemShut {NoStop}%
\bibitem [{\citenamefont {K\"ubel}\ \emph {et~al.}(2019)\citenamefont
  {K\"ubel}, \citenamefont {Katsoulis}, \citenamefont {Dube}, \citenamefont
  {Naumov}, \citenamefont {Villeneuve}, \citenamefont {Corkum}, \citenamefont
  {Staudte},\ and\ \citenamefont {Emmanouilidou}}]{Kubel19}%
  \BibitemOpen
  \bibfield  {author} {\bibinfo {author} {\bibfnamefont {M.}~\bibnamefont
  {K\"ubel}}, \bibinfo {author} {\bibfnamefont {G.~P.}\ \bibnamefont
  {Katsoulis}}, \bibinfo {author} {\bibfnamefont {Z.}~\bibnamefont {Dube}},
  \bibinfo {author} {\bibfnamefont {A.~Y.}\ \bibnamefont {Naumov}}, \bibinfo
  {author} {\bibfnamefont {D.~M.}\ \bibnamefont {Villeneuve}}, \bibinfo
  {author} {\bibfnamefont {P.~B.}\ \bibnamefont {Corkum}}, \bibinfo {author}
  {\bibfnamefont {A.}~\bibnamefont {Staudte}}, \ and\ \bibinfo {author}
  {\bibfnamefont {A.}~\bibnamefont {Emmanouilidou}},\ }\href {\doibase
  10.1103/PhysRevA.100.043410} {\bibfield  {journal} {\bibinfo  {journal}
  {Phys. Rev. A}\ }\textbf {\bibinfo {volume} {100}},\ \bibinfo {pages}
  {043410} (\bibinfo {year} {2019})}\BibitemShut {NoStop}%
\bibitem [{\citenamefont {Chen}\ \emph {et~al.}(2010)\citenamefont {Chen},
  \citenamefont {Liang},\ and\ \citenamefont {Lin}}]{Chen10}%
  \BibitemOpen
  \bibfield  {author} {\bibinfo {author} {\bibfnamefont {Z.}~\bibnamefont
  {Chen}}, \bibinfo {author} {\bibfnamefont {Y.}~\bibnamefont {Liang}}, \ and\
  \bibinfo {author} {\bibfnamefont {C.~D.}\ \bibnamefont {Lin}},\ }\href
  {\doibase 10.1103/PhysRevA.82.063417} {\bibfield  {journal} {\bibinfo
  {journal} {Phys. Rev. A}\ }\textbf {\bibinfo {volume} {82}},\ \bibinfo
  {pages} {063417} (\bibinfo {year} {2010})}\BibitemShut {NoStop}%
\bibitem [{\citenamefont {Chen}\ \emph {et~al.}(2019)\citenamefont {Chen},
  \citenamefont {Wang}, \citenamefont {Morishita}, \citenamefont {Hao},
  \citenamefont {Chen}, \citenamefont {Zatsarinny},\ and\ \citenamefont
  {Bartschat}}]{Chen19}%
  \BibitemOpen
  \bibfield  {author} {\bibinfo {author} {\bibfnamefont {Z.}~\bibnamefont
  {Chen}}, \bibinfo {author} {\bibfnamefont {Y.}~\bibnamefont {Wang}}, \bibinfo
  {author} {\bibfnamefont {T.}~\bibnamefont {Morishita}}, \bibinfo {author}
  {\bibfnamefont {X.}~\bibnamefont {Hao}}, \bibinfo {author} {\bibfnamefont
  {J.}~\bibnamefont {Chen}}, \bibinfo {author} {\bibfnamefont {O.}~\bibnamefont
  {Zatsarinny}}, \ and\ \bibinfo {author} {\bibfnamefont {K.}~\bibnamefont
  {Bartschat}},\ }\href {\doibase 10.1103/PhysRevA.100.023405} {\bibfield
  {journal} {\bibinfo  {journal} {Phys. Rev. A}\ }\textbf {\bibinfo {volume}
  {100}},\ \bibinfo {pages} {023405} (\bibinfo {year} {2019})}\BibitemShut
  {NoStop}%
\bibitem [{\citenamefont {Maxwell}\ and\ \citenamefont
  {Faria}(2016)}]{Maxwell16}%
  \BibitemOpen
  \bibfield  {author} {\bibinfo {author} {\bibfnamefont {A.~S.}\ \bibnamefont
  {Maxwell}}\ and\ \bibinfo {author} {\bibfnamefont {C.~F. d.~M.}\ \bibnamefont
  {Faria}},\ }\href {\doibase 10.1103/PhysRevLett.116.143001} {\bibfield
  {journal} {\bibinfo  {journal} {Phys. Rev. Lett.}\ }\textbf {\bibinfo
  {volume} {116}},\ \bibinfo {pages} {143001} (\bibinfo {year}
  {2016})}\BibitemShut {NoStop}%
\bibitem [{\citenamefont {Winney}\ \emph {et~al.}(2017)\citenamefont {Winney},
  \citenamefont {Lee}, \citenamefont {Lin}, \citenamefont {Liao}, \citenamefont
  {Adhikari}, \citenamefont {Basnayake}, \citenamefont {Schlegel},\ and\
  \citenamefont {Li}}]{Winney17}%
  \BibitemOpen
  \bibfield  {author} {\bibinfo {author} {\bibfnamefont {A.~H.}\ \bibnamefont
  {Winney}}, \bibinfo {author} {\bibfnamefont {S.~K.}\ \bibnamefont {Lee}},
  \bibinfo {author} {\bibfnamefont {Y.~F.}\ \bibnamefont {Lin}}, \bibinfo
  {author} {\bibfnamefont {Q.}~\bibnamefont {Liao}}, \bibinfo {author}
  {\bibfnamefont {P.}~\bibnamefont {Adhikari}}, \bibinfo {author}
  {\bibfnamefont {G.}~\bibnamefont {Basnayake}}, \bibinfo {author}
  {\bibfnamefont {H.~B.}\ \bibnamefont {Schlegel}}, \ and\ \bibinfo {author}
  {\bibfnamefont {W.}~\bibnamefont {Li}},\ }\href {\doibase
  10.1103/PhysRevLett.119.123201} {\bibfield  {journal} {\bibinfo  {journal}
  {Phys. Rev. Lett.}\ }\textbf {\bibinfo {volume} {119}},\ \bibinfo {pages}
  {123201} (\bibinfo {year} {2017})}\BibitemShut {NoStop}%
\bibitem [{\citenamefont {Winney}\ \emph {et~al.}(2018)\citenamefont {Winney},
  \citenamefont {Basnayake}, \citenamefont {Debrah}, \citenamefont {Lin},
  \citenamefont {Lee}, \citenamefont {Hoerner}, \citenamefont {Liao},
  \citenamefont {Schlegel},\ and\ \citenamefont {Li}}]{Winney18}%
  \BibitemOpen
  \bibfield  {author} {\bibinfo {author} {\bibfnamefont {A.~H.}\ \bibnamefont
  {Winney}}, \bibinfo {author} {\bibfnamefont {G.}~\bibnamefont {Basnayake}},
  \bibinfo {author} {\bibfnamefont {D.~A.}\ \bibnamefont {Debrah}}, \bibinfo
  {author} {\bibfnamefont {Y.~F.}\ \bibnamefont {Lin}}, \bibinfo {author}
  {\bibfnamefont {S.~K.}\ \bibnamefont {Lee}}, \bibinfo {author} {\bibfnamefont
  {P.}~\bibnamefont {Hoerner}}, \bibinfo {author} {\bibfnamefont
  {Q.}~\bibnamefont {Liao}}, \bibinfo {author} {\bibfnamefont {H.~B.}\
  \bibnamefont {Schlegel}}, \ and\ \bibinfo {author} {\bibfnamefont
  {W.}~\bibnamefont {Li}},\ }\href {\doibase 10.1021/acs.jpclett.8b00028}
  {\bibfield  {journal} {\bibinfo  {journal} {J. Phys. Chem. Lett.}\ }\textbf
  {\bibinfo {volume} {9}},\ \bibinfo {pages} {2539} (\bibinfo {year}
  {2018})}\BibitemShut {NoStop}%
\bibitem [{\citenamefont {Zhong}\ \emph {et~al.}(2020)\citenamefont {Zhong},
  \citenamefont {Vinbladh}, \citenamefont {Busto}, \citenamefont {Squibb},
  \citenamefont {Isinger}, \citenamefont {Neori{\v c}i{\'c}}, \citenamefont
  {Laurell}, \citenamefont {Weissenbilder}, \citenamefont {Arnold},
  \citenamefont {Feifel}, \citenamefont {Dahlstr{\"o}m}, \citenamefont
  {Wendin}, \citenamefont {Gisselbrecht}, \citenamefont {Lindroth},\ and\
  \citenamefont {L'Huillier}}]{Zhong20}%
  \BibitemOpen
  \bibfield  {author} {\bibinfo {author} {\bibfnamefont {S.}~\bibnamefont
  {Zhong}}, \bibinfo {author} {\bibfnamefont {J.}~\bibnamefont {Vinbladh}},
  \bibinfo {author} {\bibfnamefont {D.}~\bibnamefont {Busto}}, \bibinfo
  {author} {\bibfnamefont {R.~J.}\ \bibnamefont {Squibb}}, \bibinfo {author}
  {\bibfnamefont {M.}~\bibnamefont {Isinger}}, \bibinfo {author} {\bibfnamefont
  {L.}~\bibnamefont {Neori{\v c}i{\'c}}}, \bibinfo {author} {\bibfnamefont
  {H.}~\bibnamefont {Laurell}}, \bibinfo {author} {\bibfnamefont
  {R.}~\bibnamefont {Weissenbilder}}, \bibinfo {author} {\bibfnamefont {C.~L.}\
  \bibnamefont {Arnold}}, \bibinfo {author} {\bibfnamefont {R.}~\bibnamefont
  {Feifel}}, \bibinfo {author} {\bibfnamefont {J.~M.}\ \bibnamefont
  {Dahlstr{\"o}m}}, \bibinfo {author} {\bibfnamefont {G.}~\bibnamefont
  {Wendin}}, \bibinfo {author} {\bibfnamefont {M.}~\bibnamefont
  {Gisselbrecht}}, \bibinfo {author} {\bibfnamefont {E.}~\bibnamefont
  {Lindroth}}, \ and\ \bibinfo {author} {\bibfnamefont {A.}~\bibnamefont
  {L'Huillier}},\ }\href {\doibase 10.1038/s41467-020-18847-1} {\bibfield
  {journal} {\bibinfo  {journal} {Nat. Commun.}\ }\textbf {\bibinfo {volume}
  {11}},\ \bibinfo {pages} {5042} (\bibinfo {year} {2020})},\ \Eprint
  {http://arxiv.org/abs/2005.12008} {arXiv:2005.12008 [physics.atom-ph]}
  \BibitemShut {NoStop}%
\bibitem [{\citenamefont {Mikaelsson}\ \emph {et~al.}(2020)\citenamefont
  {Mikaelsson}, \citenamefont {Vogelsang}, \citenamefont {Guo}, \citenamefont
  {Sytcevich}, \citenamefont {Viotti}, \citenamefont {Langer}, \citenamefont
  {Cheng}, \citenamefont {Nandi}, \citenamefont {Jin}, \citenamefont
  {Olofsson}, \citenamefont {Weissenbilder}, \citenamefont {Mauritsson},
  \citenamefont {L'Huillier}, \citenamefont {Gisselbrecht},\ and\ \citenamefont
  {Arnold}}]{Mikaelsson20}%
  \BibitemOpen
  \bibfield  {author} {\bibinfo {author} {\bibfnamefont {S.}~\bibnamefont
  {Mikaelsson}}, \bibinfo {author} {\bibfnamefont {J.}~\bibnamefont
  {Vogelsang}}, \bibinfo {author} {\bibfnamefont {C.}~\bibnamefont {Guo}},
  \bibinfo {author} {\bibfnamefont {I.}~\bibnamefont {Sytcevich}}, \bibinfo
  {author} {\bibfnamefont {A.-L.}\ \bibnamefont {Viotti}}, \bibinfo {author}
  {\bibfnamefont {F.}~\bibnamefont {Langer}}, \bibinfo {author} {\bibfnamefont
  {Y.-C.}\ \bibnamefont {Cheng}}, \bibinfo {author} {\bibfnamefont
  {S.}~\bibnamefont {Nandi}}, \bibinfo {author} {\bibfnamefont
  {W.}~\bibnamefont {Jin}}, \bibinfo {author} {\bibfnamefont {A.}~\bibnamefont
  {Olofsson}}, \bibinfo {author} {\bibfnamefont {R.}~\bibnamefont
  {Weissenbilder}}, \bibinfo {author} {\bibfnamefont {J.}~\bibnamefont
  {Mauritsson}}, \bibinfo {author} {\bibfnamefont {A.}~\bibnamefont
  {L'Huillier}}, \bibinfo {author} {\bibfnamefont {M.}~\bibnamefont
  {Gisselbrecht}}, \ and\ \bibinfo {author} {\bibfnamefont {C.~L.}\
  \bibnamefont {Arnold}},\ }\href {\doibase 10.1515/nanoph-2020-0424}
  {\bibfield  {journal} {\bibinfo  {journal} {Nanophotonics}\ }\textbf
  {\bibinfo {volume} {10}},\ \bibinfo {pages} {117} (\bibinfo {year}
  {2020})}\BibitemShut {NoStop}%
\bibitem [{\citenamefont {Bergues}\ \emph {et~al.}(2012)\citenamefont
  {Bergues}, \citenamefont {K{\"u}bel}, \citenamefont {Johnson}, \citenamefont
  {Fischer}, \citenamefont {Camus}, \citenamefont {Betsch}, \citenamefont
  {Herrwerth}, \citenamefont {Senftleben}, \citenamefont {Sayler},
  \citenamefont {Rathje}, \citenamefont {Pfeifer}, \citenamefont {Ben-Itzhak},
  \citenamefont {Jones}, \citenamefont {Paulus}, \citenamefont {Krausz},
  \citenamefont {Moshammer}, \citenamefont {Ullrich},\ and\ \citenamefont
  {Kling}}]{Bergues12}%
  \BibitemOpen
  \bibfield  {author} {\bibinfo {author} {\bibfnamefont {B.}~\bibnamefont
  {Bergues}}, \bibinfo {author} {\bibfnamefont {M.}~\bibnamefont {K{\"u}bel}},
  \bibinfo {author} {\bibfnamefont {N.~G.}\ \bibnamefont {Johnson}}, \bibinfo
  {author} {\bibfnamefont {B.}~\bibnamefont {Fischer}}, \bibinfo {author}
  {\bibfnamefont {N.}~\bibnamefont {Camus}}, \bibinfo {author} {\bibfnamefont
  {K.~J.}\ \bibnamefont {Betsch}}, \bibinfo {author} {\bibfnamefont
  {O.}~\bibnamefont {Herrwerth}}, \bibinfo {author} {\bibfnamefont
  {A.}~\bibnamefont {Senftleben}}, \bibinfo {author} {\bibfnamefont {A.~M.}\
  \bibnamefont {Sayler}}, \bibinfo {author} {\bibfnamefont {T.}~\bibnamefont
  {Rathje}}, \bibinfo {author} {\bibfnamefont {T.}~\bibnamefont {Pfeifer}},
  \bibinfo {author} {\bibfnamefont {I.}~\bibnamefont {Ben-Itzhak}}, \bibinfo
  {author} {\bibfnamefont {R.~R.}\ \bibnamefont {Jones}}, \bibinfo {author}
  {\bibfnamefont {G.~G.}\ \bibnamefont {Paulus}}, \bibinfo {author}
  {\bibfnamefont {F.}~\bibnamefont {Krausz}}, \bibinfo {author} {\bibfnamefont
  {R.}~\bibnamefont {Moshammer}}, \bibinfo {author} {\bibfnamefont
  {J.}~\bibnamefont {Ullrich}}, \ and\ \bibinfo {author} {\bibfnamefont
  {M.~F.}\ \bibnamefont {Kling}},\ }\href@noop {} {\bibfield  {journal}
  {\bibinfo  {journal} {Nat. Commun.}\ }\textbf {\bibinfo {volume} {3}},\
  \bibinfo {pages} {813} (\bibinfo {year} {2012})}\BibitemShut {NoStop}%
\bibitem [{\citenamefont {Henrichs}\ \emph {et~al.}(2018)\citenamefont
  {Henrichs}, \citenamefont {Eckart}, \citenamefont {Hartung}, \citenamefont
  {Trabert}, \citenamefont {Fehre}, \citenamefont {Rist}, \citenamefont {Sann},
  \citenamefont {Pitzer}, \citenamefont {Richter}, \citenamefont {Kang},
  \citenamefont {Sch{\"o}ffler}, \citenamefont {Kunitski}, \citenamefont
  {Jahnke},\ and\ \citenamefont {D{\"o}rner}}]{Henrichs18}%
  \BibitemOpen
  \bibfield  {author} {\bibinfo {author} {\bibfnamefont {K.}~\bibnamefont
  {Henrichs}}, \bibinfo {author} {\bibfnamefont {S.}~\bibnamefont {Eckart}},
  \bibinfo {author} {\bibfnamefont {A.}~\bibnamefont {Hartung}}, \bibinfo
  {author} {\bibfnamefont {D.}~\bibnamefont {Trabert}}, \bibinfo {author}
  {\bibfnamefont {K.}~\bibnamefont {Fehre}}, \bibinfo {author} {\bibfnamefont
  {J.}~\bibnamefont {Rist}}, \bibinfo {author} {\bibfnamefont {H.}~\bibnamefont
  {Sann}}, \bibinfo {author} {\bibfnamefont {M.}~\bibnamefont {Pitzer}},
  \bibinfo {author} {\bibfnamefont {M.}~\bibnamefont {Richter}}, \bibinfo
  {author} {\bibfnamefont {H.}~\bibnamefont {Kang}}, \bibinfo {author}
  {\bibfnamefont {M.~S.}\ \bibnamefont {Sch{\"o}ffler}}, \bibinfo {author}
  {\bibfnamefont {M.}~\bibnamefont {Kunitski}}, \bibinfo {author}
  {\bibfnamefont {T.}~\bibnamefont {Jahnke}}, \ and\ \bibinfo {author}
  {\bibfnamefont {R.}~\bibnamefont {D{\"o}rner}},\ }\href {\doibase
  10.1103/PhysRevA.98.043405} {\bibfield  {journal} {\bibinfo  {journal} {Phys.
  Rev. A}\ }\textbf {\bibinfo {volume} {98}},\ \bibinfo {pages} {043405}
  (\bibinfo {year} {2018})}\BibitemShut {NoStop}%
\bibitem [{\citenamefont {Larimian}\ \emph {et~al.}(2020)\citenamefont
  {Larimian}, \citenamefont {Erattupuzha}, \citenamefont {Baltu{\v s}ka},
  \citenamefont {Kitzler-Zeiler},\ and\ \citenamefont {Xie}}]{Larimian20}%
  \BibitemOpen
  \bibfield  {author} {\bibinfo {author} {\bibfnamefont {S.}~\bibnamefont
  {Larimian}}, \bibinfo {author} {\bibfnamefont {S.}~\bibnamefont
  {Erattupuzha}}, \bibinfo {author} {\bibfnamefont {A.}~\bibnamefont {Baltu{\v
  s}ka}}, \bibinfo {author} {\bibfnamefont {M.}~\bibnamefont {Kitzler-Zeiler}},
  \ and\ \bibinfo {author} {\bibfnamefont {X.}~\bibnamefont {Xie}},\ }\href
  {\doibase 10.1103/PhysRevResearch.2.013021} {\bibfield  {journal} {\bibinfo
  {journal} {Phys. Rev. Research}\ }\textbf {\bibinfo {volume} {2}},\ \bibinfo
  {pages} {013021} (\bibinfo {year} {2020})}\BibitemShut {NoStop}%
\bibitem [{\citenamefont {Grundmann}\ \emph {et~al.}(2020)\citenamefont
  {Grundmann}, \citenamefont {Serov}, \citenamefont {Trinter}, \citenamefont
  {Fehre}, \citenamefont {Strenger}, \citenamefont {Pier}, \citenamefont
  {Kircher}, \citenamefont {Trabert}, \citenamefont {Weller}, \citenamefont
  {Rist}, \citenamefont {Kaiser}, \citenamefont {Bray}, \citenamefont
  {Schmidt}, \citenamefont {Williams}, \citenamefont {Jahnke}, \citenamefont
  {D{\"o}rner}, \citenamefont {Sch{\"o}ffler},\ and\ \citenamefont
  {Kheifets}}]{Grundmann20}%
  \BibitemOpen
  \bibfield  {author} {\bibinfo {author} {\bibfnamefont {S.}~\bibnamefont
  {Grundmann}}, \bibinfo {author} {\bibfnamefont {V.~V.}\ \bibnamefont
  {Serov}}, \bibinfo {author} {\bibfnamefont {F.}~\bibnamefont {Trinter}},
  \bibinfo {author} {\bibfnamefont {K.}~\bibnamefont {Fehre}}, \bibinfo
  {author} {\bibfnamefont {N.}~\bibnamefont {Strenger}}, \bibinfo {author}
  {\bibfnamefont {A.}~\bibnamefont {Pier}}, \bibinfo {author} {\bibfnamefont
  {M.}~\bibnamefont {Kircher}}, \bibinfo {author} {\bibfnamefont
  {D.}~\bibnamefont {Trabert}}, \bibinfo {author} {\bibfnamefont
  {M.}~\bibnamefont {Weller}}, \bibinfo {author} {\bibfnamefont
  {J.}~\bibnamefont {Rist}}, \bibinfo {author} {\bibfnamefont {L.}~\bibnamefont
  {Kaiser}}, \bibinfo {author} {\bibfnamefont {A.~W.}\ \bibnamefont {Bray}},
  \bibinfo {author} {\bibfnamefont {L.~P.~H.}\ \bibnamefont {Schmidt}},
  \bibinfo {author} {\bibfnamefont {J.~B.}\ \bibnamefont {Williams}}, \bibinfo
  {author} {\bibfnamefont {T.}~\bibnamefont {Jahnke}}, \bibinfo {author}
  {\bibfnamefont {R.}~\bibnamefont {D{\"o}rner}}, \bibinfo {author}
  {\bibfnamefont {M.~S.}\ \bibnamefont {Sch{\"o}ffler}}, \ and\ \bibinfo
  {author} {\bibfnamefont {A.~S.}\ \bibnamefont {Kheifets}},\ }\href {\doibase
  10.1103/PhysRevResearch.2.033080} {\bibfield  {journal} {\bibinfo  {journal}
  {Phys. Rev. Research}\ }\textbf {\bibinfo {volume} {2}},\ \bibinfo {pages}
  {033080} (\bibinfo {year} {2020})}\BibitemShut {NoStop}%
\bibitem [{\citenamefont {Schulz}\ \emph {et~al.}(2000)\citenamefont {Schulz},
  \citenamefont {Moshammer}, \citenamefont {Schmitt}, \citenamefont {Kollmus},
  \citenamefont {Mann}, \citenamefont {Hagmann}, \citenamefont {Olson},\ and\
  \citenamefont {Ullrich}}]{Schulz00}%
  \BibitemOpen
  \bibfield  {author} {\bibinfo {author} {\bibfnamefont {M.}~\bibnamefont
  {Schulz}}, \bibinfo {author} {\bibfnamefont {R.}~\bibnamefont {Moshammer}},
  \bibinfo {author} {\bibfnamefont {W.}~\bibnamefont {Schmitt}}, \bibinfo
  {author} {\bibfnamefont {H.}~\bibnamefont {Kollmus}}, \bibinfo {author}
  {\bibfnamefont {R.}~\bibnamefont {Mann}}, \bibinfo {author} {\bibfnamefont
  {S.}~\bibnamefont {Hagmann}}, \bibinfo {author} {\bibfnamefont {R.~E.}\
  \bibnamefont {Olson}}, \ and\ \bibinfo {author} {\bibfnamefont
  {J.}~\bibnamefont {Ullrich}},\ }\href {\doibase 10.1103/PhysRevA.61.022703}
  {\bibfield  {journal} {\bibinfo  {journal} {Phys. Rev. A}\ }\textbf {\bibinfo
  {volume} {61}},\ \bibinfo {pages} {022703} (\bibinfo {year}
  {2000})}\BibitemShut {NoStop}%
\bibitem [{\citenamefont {Efimov}\ \emph {et~al.}(2019)\citenamefont {Efimov},
  \citenamefont {Prauzner-Bechcicki}, \citenamefont {Thiede}, \citenamefont
  {Eckhardt},\ and\ \citenamefont {Zakrzewski}}]{Efimov19}%
  \BibitemOpen
  \bibfield  {author} {\bibinfo {author} {\bibfnamefont {D.~K.}\ \bibnamefont
  {Efimov}}, \bibinfo {author} {\bibfnamefont {J.~S.}\ \bibnamefont
  {Prauzner-Bechcicki}}, \bibinfo {author} {\bibfnamefont {J.~H.}\ \bibnamefont
  {Thiede}}, \bibinfo {author} {\bibfnamefont {B.}~\bibnamefont {Eckhardt}}, \
  and\ \bibinfo {author} {\bibfnamefont {J.}~\bibnamefont {Zakrzewski}},\
  }\href {\doibase 10.1103/PhysRevA.100.063408} {\bibfield  {journal} {\bibinfo
   {journal} {Phys. Rev. A}\ }\textbf {\bibinfo {volume} {100}},\ \bibinfo
  {pages} {063408} (\bibinfo {year} {2019})}\BibitemShut {NoStop}%
\bibitem [{\citenamefont {Prauzner-Bechcicki}\ \emph
  {et~al.}(2021)\citenamefont {Prauzner-Bechcicki}, \citenamefont {Efimov},
  \citenamefont {Mandrysz},\ and\ \citenamefont
  {Zakrzewski}}]{Prauzner-Bechcicki21}%
  \BibitemOpen
  \bibfield  {author} {\bibinfo {author} {\bibfnamefont {J.~S.}\ \bibnamefont
  {Prauzner-Bechcicki}}, \bibinfo {author} {\bibfnamefont {D.~K.}\ \bibnamefont
  {Efimov}}, \bibinfo {author} {\bibfnamefont {M.}~\bibnamefont {Mandrysz}}, \
  and\ \bibinfo {author} {\bibfnamefont {J.}~\bibnamefont {Zakrzewski}},\
  }\href@noop {} {\bibfield  {journal} {\bibinfo  {journal} {Journal of Physics
  B: Atomic, Molecular and Optical Physics}\ }\textbf {\bibinfo {volume}
  {accepted}} (\bibinfo {year} {2021})},\ \Eprint
  {http://arxiv.org/abs/2102.06466} {arXiv:2102.06466 [physics.atom-ph]}
  \BibitemShut {NoStop}%
\bibitem [{\citenamefont {Parker}\ \emph {et~al.}(1998)\citenamefont {Parker},
  \citenamefont {Smyth},\ and\ \citenamefont {Taylor}}]{Parker98}%
  \BibitemOpen
  \bibfield  {author} {\bibinfo {author} {\bibfnamefont {J.~S.}\ \bibnamefont
  {Parker}}, \bibinfo {author} {\bibfnamefont {E.~S.}\ \bibnamefont {Smyth}}, \
  and\ \bibinfo {author} {\bibfnamefont {K.~T.}\ \bibnamefont {Taylor}},\
  }\href@noop {} {\bibfield  {journal} {\bibinfo  {journal} {J. Phys. B-At.
  Mol. Opt.}\ }\textbf {\bibinfo {volume} {31}},\ \bibinfo {pages} {L571}
  (\bibinfo {year} {1998})}\BibitemShut {NoStop}%
\bibitem [{\citenamefont {Parker}\ \emph {et~al.}(2000)\citenamefont {Parker},
  \citenamefont {Glass}, \citenamefont {Moore}, \citenamefont {Smyth},
  \citenamefont {Taylor},\ and\ \citenamefont {Burke}}]{Parker00}%
  \BibitemOpen
  \bibfield  {author} {\bibinfo {author} {\bibfnamefont {J.~S.}\ \bibnamefont
  {Parker}}, \bibinfo {author} {\bibfnamefont {D.}~\bibnamefont {Glass}},
  \bibinfo {author} {\bibfnamefont {L.~R.}\ \bibnamefont {Moore}}, \bibinfo
  {author} {\bibfnamefont {E.~S.}\ \bibnamefont {Smyth}}, \bibinfo {author}
  {\bibfnamefont {K.}~\bibnamefont {Taylor}}, \ and\ \bibinfo {author}
  {\bibfnamefont {P.}~\bibnamefont {Burke}},\ }\href@noop {} {\bibfield
  {journal} {\bibinfo  {journal} {J. Phys. B-At. Mol. Opt.}\ }\textbf {\bibinfo
  {volume} {33}},\ \bibinfo {pages} {L239} (\bibinfo {year}
  {2000})}\BibitemShut {NoStop}%
\bibitem [{\citenamefont {Emmanouilidou}\ \emph {et~al.}(2008)\citenamefont
  {Emmanouilidou}, \citenamefont {Wang},\ and\ \citenamefont
  {Rost}}]{Emmanouilidou08b}%
  \BibitemOpen
  \bibfield  {author} {\bibinfo {author} {\bibfnamefont {A.}~\bibnamefont
  {Emmanouilidou}}, \bibinfo {author} {\bibfnamefont {P.}~\bibnamefont {Wang}},
  \ and\ \bibinfo {author} {\bibfnamefont {J.~M.}\ \bibnamefont {Rost}},\
  }\href {\doibase 10.1103/PhysRevLett.100.063002} {\bibfield  {journal}
  {\bibinfo  {journal} {Phys. Rev. Lett.}\ }\textbf {\bibinfo {volume} {100}},\
  \bibinfo {pages} {063002} (\bibinfo {year} {2008})}\BibitemShut {NoStop}%
\bibitem [{\citenamefont {Peters}\ \emph {et~al.}(2020)\citenamefont {Peters},
  \citenamefont {Majety},\ and\ \citenamefont {Emmanouilidou}}]{Peters20}%
  \BibitemOpen
  \bibfield  {author} {\bibinfo {author} {\bibfnamefont {M.~B.}\ \bibnamefont
  {Peters}}, \bibinfo {author} {\bibfnamefont {V.~P.}\ \bibnamefont {Majety}},
  \ and\ \bibinfo {author} {\bibfnamefont {A.}~\bibnamefont {Emmanouilidou}},\
  }\href {http://arxiv.org/abs/2010.16216} {\bibfield  {journal} {\bibinfo
  {journal} {arXiv preprint arXiv:2010.16216}\ } (\bibinfo {year} {2020})},\
  \Eprint {http://arxiv.org/abs/2010.16216} {arXiv:2010.16216
  [physics.atom-ph]} \BibitemShut {NoStop}%
\bibitem [{\citenamefont {Ho}\ and\ \citenamefont
  {Eberly}(2006)}]{ho2006plane}%
  \BibitemOpen
  \bibfield  {author} {\bibinfo {author} {\bibfnamefont {P.~J.}\ \bibnamefont
  {Ho}}\ and\ \bibinfo {author} {\bibfnamefont {J.}~\bibnamefont {Eberly}},\
  }\href@noop {} {\bibfield  {journal} {\bibinfo  {journal} {Phys. Rev. Lett.}\
  }\textbf {\bibinfo {volume} {97}},\ \bibinfo {pages} {083001} (\bibinfo
  {year} {2006})}\BibitemShut {NoStop}%
\bibitem [{\citenamefont {Ho}\ and\ \citenamefont
  {Eberly}(2007)}]{ho2007argon}%
  \BibitemOpen
  \bibfield  {author} {\bibinfo {author} {\bibfnamefont {P.~J.}\ \bibnamefont
  {Ho}}\ and\ \bibinfo {author} {\bibfnamefont {J.}~\bibnamefont {Eberly}},\
  }\href@noop {} {\bibfield  {journal} {\bibinfo  {journal} {Optics Express}\
  }\textbf {\bibinfo {volume} {15}},\ \bibinfo {pages} {1845} (\bibinfo {year}
  {2007})}\BibitemShut {NoStop}%
\bibitem [{\citenamefont {Guo}\ and\ \citenamefont {Liu}(2008)}]{Guo08}%
  \BibitemOpen
  \bibfield  {author} {\bibinfo {author} {\bibfnamefont {J.}~\bibnamefont
  {Guo}}\ and\ \bibinfo {author} {\bibfnamefont {X.-s.}\ \bibnamefont {Liu}},\
  }\href {\doibase 10.1103/PhysRevA.78.013401} {\bibfield  {journal} {\bibinfo
  {journal} {Phys. Rev. A}\ }\textbf {\bibinfo {volume} {78}},\ \bibinfo
  {pages} {013401} (\bibinfo {year} {2008})}\BibitemShut {NoStop}%
\bibitem [{\citenamefont {Yuan}\ \emph {et~al.}(2019)\citenamefont {Yuan},
  \citenamefont {Ye}, \citenamefont {Gu}, \citenamefont {Liu},\ and\
  \citenamefont {Fu}}]{Yuan19}%
  \BibitemOpen
  \bibfield  {author} {\bibinfo {author} {\bibfnamefont {Z.-Q.}\ \bibnamefont
  {Yuan}}, \bibinfo {author} {\bibfnamefont {D.-F.}\ \bibnamefont {Ye}},
  \bibinfo {author} {\bibfnamefont {Y.-Q.}\ \bibnamefont {Gu}}, \bibinfo
  {author} {\bibfnamefont {J.}~\bibnamefont {Liu}}, \ and\ \bibinfo {author}
  {\bibfnamefont {L.-B.}\ \bibnamefont {Fu}},\ }\href {\doibase
  10.1364/OE.27.003180} {\bibfield  {journal} {\bibinfo  {journal} {Opt.
  Express}\ }\textbf {\bibinfo {volume} {27}},\ \bibinfo {pages} {3180}
  (\bibinfo {year} {2019})}\BibitemShut {NoStop}%
\bibitem [{\citenamefont {Sato}\ \emph {et~al.}(2016)\citenamefont {Sato},
  \citenamefont {Ishikawa}, \citenamefont {B\ifmmode~\check{r}\else
  \v{r}\fi{}ezinov\'a}, \citenamefont {Lackner}, \citenamefont {Nagele},\ and\
  \citenamefont {Burgd\"orfer}}]{Sato16}%
  \BibitemOpen
  \bibfield  {author} {\bibinfo {author} {\bibfnamefont {T.}~\bibnamefont
  {Sato}}, \bibinfo {author} {\bibfnamefont {K.~L.}\ \bibnamefont {Ishikawa}},
  \bibinfo {author} {\bibfnamefont {I.}~\bibnamefont {B\ifmmode~\check{r}\else
  \v{r}\fi{}ezinov\'a}}, \bibinfo {author} {\bibfnamefont {F.}~\bibnamefont
  {Lackner}}, \bibinfo {author} {\bibfnamefont {S.}~\bibnamefont {Nagele}}, \
  and\ \bibinfo {author} {\bibfnamefont {J.}~\bibnamefont {Burgd\"orfer}},\
  }\href {\doibase 10.1103/PhysRevA.94.023405} {\bibfield  {journal} {\bibinfo
  {journal} {Phys. Rev. A}\ }\textbf {\bibinfo {volume} {94}},\ \bibinfo
  {pages} {023405} (\bibinfo {year} {2016})}\BibitemShut {NoStop}%
\bibitem [{\citenamefont {Thiede}\ \emph {et~al.}(2018)\citenamefont {Thiede},
  \citenamefont {Eckhardt}, \citenamefont {Efimov}, \citenamefont
  {Prauzner-Bechcicki},\ and\ \citenamefont {Zakrzewski}}]{Thiede18}%
  \BibitemOpen
  \bibfield  {author} {\bibinfo {author} {\bibfnamefont {J.~H.}\ \bibnamefont
  {Thiede}}, \bibinfo {author} {\bibfnamefont {B.}~\bibnamefont {Eckhardt}},
  \bibinfo {author} {\bibfnamefont {D.~K.}\ \bibnamefont {Efimov}}, \bibinfo
  {author} {\bibfnamefont {J.~S.}\ \bibnamefont {Prauzner-Bechcicki}}, \ and\
  \bibinfo {author} {\bibfnamefont {J.}~\bibnamefont {Zakrzewski}},\ }\href
  {\doibase 10.1103/PhysRevA.98.031401} {\bibfield  {journal} {\bibinfo
  {journal} {Phys. Rev. A}\ }\textbf {\bibinfo {volume} {98}},\ \bibinfo
  {pages} {031401} (\bibinfo {year} {2018})}\BibitemShut {NoStop}%
\bibitem [{\citenamefont {Efimov}\ \emph {et~al.}(2020)\citenamefont {Efimov},
  \citenamefont {Prauzner-Bechcicki},\ and\ \citenamefont
  {Zakrzewski}}]{Efimov20}%
  \BibitemOpen
  \bibfield  {author} {\bibinfo {author} {\bibfnamefont {D.~K.}\ \bibnamefont
  {Efimov}}, \bibinfo {author} {\bibfnamefont {J.~S.}\ \bibnamefont
  {Prauzner-Bechcicki}}, \ and\ \bibinfo {author} {\bibfnamefont
  {J.}~\bibnamefont {Zakrzewski}},\ }\href {\doibase
  10.1103/PhysRevA.101.063402} {\bibfield  {journal} {\bibinfo  {journal}
  {Phys. Rev. A}\ }\textbf {\bibinfo {volume} {101}},\ \bibinfo {pages}
  {063402} (\bibinfo {year} {2020})}\BibitemShut {NoStop}%
\bibitem [{\citenamefont {Grobe}\ and\ \citenamefont {Eberly}(1992)}]{Grobe92}%
  \BibitemOpen
  \bibfield  {author} {\bibinfo {author} {\bibfnamefont {R.}~\bibnamefont
  {Grobe}}\ and\ \bibinfo {author} {\bibfnamefont {J.~H.}\ \bibnamefont
  {Eberly}},\ }\href {\doibase 10.1103/PhysRevLett.68.2905} {\bibfield
  {journal} {\bibinfo  {journal} {Phys. Rev. Lett.}\ }\textbf {\bibinfo
  {volume} {68}},\ \bibinfo {pages} {2905} (\bibinfo {year}
  {1992})}\BibitemShut {NoStop}%
\bibitem [{\citenamefont {Liu}\ \emph {et~al.}(1999)\citenamefont {Liu},
  \citenamefont {Eberly}, \citenamefont {Haan},\ and\ \citenamefont
  {Grobe}}]{Liu99}%
  \BibitemOpen
  \bibfield  {author} {\bibinfo {author} {\bibfnamefont {W.-C.}\ \bibnamefont
  {Liu}}, \bibinfo {author} {\bibfnamefont {J.~H.}\ \bibnamefont {Eberly}},
  \bibinfo {author} {\bibfnamefont {S.~L.}\ \bibnamefont {Haan}}, \ and\
  \bibinfo {author} {\bibfnamefont {R.}~\bibnamefont {Grobe}},\ }\href
  {\doibase 10.1103/PhysRevLett.83.520} {\bibfield  {journal} {\bibinfo
  {journal} {Phys. Rev. Lett.}\ }\textbf {\bibinfo {volume} {83}},\ \bibinfo
  {pages} {520} (\bibinfo {year} {1999})}\BibitemShut {NoStop}%
\bibitem [{\citenamefont {Sacha}\ and\ \citenamefont
  {Eckhardt}(2001)}]{Sacha01}%
  \BibitemOpen
  \bibfield  {author} {\bibinfo {author} {\bibfnamefont {K.}~\bibnamefont
  {Sacha}}\ and\ \bibinfo {author} {\bibfnamefont {B.}~\bibnamefont
  {Eckhardt}},\ }\href {\doibase 10.1103/PhysRevA.64.053401} {\bibfield
  {journal} {\bibinfo  {journal} {Phys. Rev. A}\ }\textbf {\bibinfo {volume}
  {64}},\ \bibinfo {pages} {053401} (\bibinfo {year} {2001})}\BibitemShut
  {NoStop}%
\bibitem [{\citenamefont {Lein}\ \emph {et~al.}(2000)\citenamefont {Lein},
  \citenamefont {Gross},\ and\ \citenamefont {Engel}}]{Lein00}%
  \BibitemOpen
  \bibfield  {author} {\bibinfo {author} {\bibfnamefont {M.}~\bibnamefont
  {Lein}}, \bibinfo {author} {\bibfnamefont {E.~K.~U.}\ \bibnamefont {Gross}},
  \ and\ \bibinfo {author} {\bibfnamefont {V.}~\bibnamefont {Engel}},\ }\href
  {\doibase 10.1103/PhysRevLett.85.4707} {\bibfield  {journal} {\bibinfo
  {journal} {Phys. Rev. Lett.}\ }\textbf {\bibinfo {volume} {85}},\ \bibinfo
  {pages} {4707} (\bibinfo {year} {2000})}\BibitemShut {NoStop}%
\bibitem [{\citenamefont {Prauzner-Bechcicki}\ \emph
  {et~al.}(2008)\citenamefont {Prauzner-Bechcicki}, \citenamefont {Sacha},
  \citenamefont {Eckhardt},\ and\ \citenamefont {Zakrzewski}}]{Prauzner08}%
  \BibitemOpen
  \bibfield  {author} {\bibinfo {author} {\bibfnamefont {J.~S.}\ \bibnamefont
  {Prauzner-Bechcicki}}, \bibinfo {author} {\bibfnamefont {K.}~\bibnamefont
  {Sacha}}, \bibinfo {author} {\bibfnamefont {B.}~\bibnamefont {Eckhardt}}, \
  and\ \bibinfo {author} {\bibfnamefont {J.}~\bibnamefont {Zakrzewski}},\
  }\href {\doibase 10.1103/PhysRevA.78.013419} {\bibfield  {journal} {\bibinfo
  {journal} {Phys. Rev. A}\ }\textbf {\bibinfo {volume} {78}},\ \bibinfo
  {pages} {013419} (\bibinfo {year} {2008})}\BibitemShut {NoStop}%
\bibitem [{\citenamefont {Ruiz}\ \emph {et~al.}(2005)\citenamefont {Ruiz},
  \citenamefont {Plaja},\ and\ \citenamefont {Roso}}]{Ruiz05}%
  \BibitemOpen
  \bibfield  {author} {\bibinfo {author} {\bibfnamefont {C.}~\bibnamefont
  {Ruiz}}, \bibinfo {author} {\bibfnamefont {L.}~\bibnamefont {Plaja}}, \ and\
  \bibinfo {author} {\bibfnamefont {L.}~\bibnamefont {Roso}},\ }\href {\doibase
  10.1103/PhysRevLett.94.063002} {\bibfield  {journal} {\bibinfo  {journal}
  {Phys. Rev. Lett.}\ }\textbf {\bibinfo {volume} {94}},\ \bibinfo {pages}
  {063002} (\bibinfo {year} {2005})}\BibitemShut {NoStop}%
\bibitem [{\citenamefont {Ruiz}\ \emph {et~al.}(2006)\citenamefont {Ruiz},
  \citenamefont {Plaja},\ and\ \citenamefont {Roso}}]{Ruiz06}%
  \BibitemOpen
  \bibfield  {author} {\bibinfo {author} {\bibfnamefont {C.}~\bibnamefont
  {Ruiz}}, \bibinfo {author} {\bibfnamefont {L.}~\bibnamefont {Plaja}}, \ and\
  \bibinfo {author} {\bibfnamefont {L.}~\bibnamefont {Roso}},\ }\href {\doibase
  10.1134/S1054660X06040116} {\bibfield  {journal} {\bibinfo  {journal} {Laser
  Physics}\ }\textbf {\bibinfo {volume} {16}},\ \bibinfo {pages} {600}
  (\bibinfo {year} {2006})}\BibitemShut {NoStop}%
\bibitem [{sup()}]{suppl}%
  \BibitemOpen
  \href@noop {} {}\bibinfo {note} {See supplementary material at [URL will be
  inserted by publisher] for details on numerical integration of TDSE as well
  as on spin-sensitive Dalitz plots.}\BibitemShut {Stop}%
\bibitem [{\citenamefont {Ciappina}\ \emph {et~al.}(2006)\citenamefont
  {Ciappina}, \citenamefont {Cravero}, \citenamefont {Schulz}, \citenamefont
  {Moshammer},\ and\ \citenamefont {Ullrich}}]{Ciappina06}%
  \BibitemOpen
  \bibfield  {author} {\bibinfo {author} {\bibfnamefont {M.~F.}\ \bibnamefont
  {Ciappina}}, \bibinfo {author} {\bibfnamefont {W.~R.}\ \bibnamefont
  {Cravero}}, \bibinfo {author} {\bibfnamefont {M.}~\bibnamefont {Schulz}},
  \bibinfo {author} {\bibfnamefont {R.}~\bibnamefont {Moshammer}}, \ and\
  \bibinfo {author} {\bibfnamefont {J.}~\bibnamefont {Ullrich}},\ }\href
  {\doibase 10.1103/PhysRevA.74.042702} {\bibfield  {journal} {\bibinfo
  {journal} {Phys. Rev. A}\ }\textbf {\bibinfo {volume} {74}},\ \bibinfo
  {pages} {042702} (\bibinfo {year} {2006})}\BibitemShut {NoStop}%
\bibitem [{\citenamefont {Staudte}\ \emph {et~al.}(2007)\citenamefont
  {Staudte}, \citenamefont {Ruiz}, \citenamefont {Sch\"offler}, \citenamefont
  {Sch\"ossler}, \citenamefont {Zeidler}, \citenamefont {Weber}, \citenamefont
  {Meckel}, \citenamefont {Villeneuve}, \citenamefont {Corkum}, \citenamefont
  {Becker},\ and\ \citenamefont {D\"orner}}]{staudte2007binary}%
  \BibitemOpen
  \bibfield  {author} {\bibinfo {author} {\bibfnamefont {A.}~\bibnamefont
  {Staudte}}, \bibinfo {author} {\bibfnamefont {C.}~\bibnamefont {Ruiz}},
  \bibinfo {author} {\bibfnamefont {M.}~\bibnamefont {Sch\"offler}}, \bibinfo
  {author} {\bibfnamefont {S.}~\bibnamefont {Sch\"ossler}}, \bibinfo {author}
  {\bibfnamefont {D.}~\bibnamefont {Zeidler}}, \bibinfo {author} {\bibfnamefont
  {T.}~\bibnamefont {Weber}}, \bibinfo {author} {\bibfnamefont
  {M.}~\bibnamefont {Meckel}}, \bibinfo {author} {\bibfnamefont {D.~M.}\
  \bibnamefont {Villeneuve}}, \bibinfo {author} {\bibfnamefont {P.~B.}\
  \bibnamefont {Corkum}}, \bibinfo {author} {\bibfnamefont {A.}~\bibnamefont
  {Becker}}, \ and\ \bibinfo {author} {\bibfnamefont {R.}~\bibnamefont
  {D\"orner}},\ }\href {\doibase 10.1103/PhysRevLett.99.263002} {\bibfield
  {journal} {\bibinfo  {journal} {Phys. Rev. Lett.}\ }\textbf {\bibinfo
  {volume} {99}},\ \bibinfo {pages} {263002} (\bibinfo {year}
  {2007})}\BibitemShut {NoStop}%
\bibitem [{\citenamefont {Rudenko}\ \emph {et~al.}(2007)\citenamefont
  {Rudenko}, \citenamefont {de~Jesus}, \citenamefont {Ergler}, \citenamefont
  {Zrost}, \citenamefont {Feuerstein}, \citenamefont {Schr\"oter},
  \citenamefont {Moshammer},\ and\ \citenamefont
  {Ullrich}}]{rudenko2007correlated}%
  \BibitemOpen
  \bibfield  {author} {\bibinfo {author} {\bibfnamefont {A.}~\bibnamefont
  {Rudenko}}, \bibinfo {author} {\bibfnamefont {V.~L.~B.}\ \bibnamefont
  {de~Jesus}}, \bibinfo {author} {\bibfnamefont {T.}~\bibnamefont {Ergler}},
  \bibinfo {author} {\bibfnamefont {K.}~\bibnamefont {Zrost}}, \bibinfo
  {author} {\bibfnamefont {B.}~\bibnamefont {Feuerstein}}, \bibinfo {author}
  {\bibfnamefont {C.~D.}\ \bibnamefont {Schr\"oter}}, \bibinfo {author}
  {\bibfnamefont {R.}~\bibnamefont {Moshammer}}, \ and\ \bibinfo {author}
  {\bibfnamefont {J.}~\bibnamefont {Ullrich}},\ }\href {\doibase
  10.1103/PhysRevLett.99.263003} {\bibfield  {journal} {\bibinfo  {journal}
  {Phys. Rev. Lett.}\ }\textbf {\bibinfo {volume} {99}},\ \bibinfo {pages}
  {263003} (\bibinfo {year} {2007})}\BibitemShut {NoStop}%
\end{thebibliography}
